%% file: main.tex
\documentclass[a4paper,11pt]{article}
\pdfoutput=1 % if your are submitting a pdflatex (i.e. if you have
\usepackage{jinst} % for details on the use of the package, please
\usepackage{subcaption}
\usepackage{bm}

\title{Measurement of the Longitudinal Diffusion of Ionization Electrons in the MicroBooNE Detector}
\collaboration{MicroBooNE Collaboration}
\begin{document}

% make sure the \maketitle command appears AFTER the affiliations!

% For signal processing paper #2 (Feb 2018), add invited authors: Mike Dolce, Veljko Radeka, Bo Yu, Ivan Caro Terrazas, Bo Yu
% For signal processing paper #1 (Jan 2018), add invited authors: Veljko Radeka, Bo Yu, Craig Thorn, Bo Yu
% December 2017: see notes in docdb #13001
% July 2017 MuCS: add Harvard, joint affiliations for Adams, Guenette, Terao
% For Noise paper (Apr 2017), add invited authors: B. Bullard, G. De Geronimo, S. Li, V. Radeka, S. Rescia, C. Thorn, B. Yu
% For MuCS paper (Mar 2017), add invited authors: L.N. Kalousis, G. Lange, R. Pelkey
% For MCS paper (Feb 2017), add invited authors: L.N. Kalousis, P. Abratenko

% Authors in alphabetical order
\author[jj]{P.~Abratenko}
\author[o]{R.~An}
\author[d]{J.~Anthony}
\author[ii]{J.~Asaadi}
\author[t,gg]{A.~Ashkenazi}
\author[mm,l]{S.~Balasubramanian}
\author[l]{B.~Baller}
\author[u]{C.~Barnes}
\author[y]{G.~Barr}
\author[s]{V.~Basque}
% ...

% Institutions in alphabetical order
% ...
\affiliation[d]{University of Cambridge, Cambridge CB3 0HE, United Kingdom}
% ...
\affiliation[l]{Fermi National Accelerator Laboratory (FNAL), Batavia, IL 60510, USA}
% ...
\affiliation[o]{Illinois Institute of Technology (IIT), Chicago, IL 60616, USA}
% ...
\affiliation[s]{The University of Manchester, Manchester M13 9PL, United Kingdom}
\affiliation[t]{Massachusetts Institute of Technology (MIT), Cambridge, MA, 02139, USA}
\affiliation[u]{University of Michigan, Ann Arbor, MI, 48109, USA}
% ...
\affiliation[y]{University of Oxford, Oxford OX1 3RH, United Kingdom}
% ...
\affiliation[gg]{Tel Aviv University, Tel Aviv, Israel, 69978}
% ...
\affiliation[ii]{University of Texas, Arlington, TX, 76019, USA}
\affiliation[jj]{Tufts University, Medford, MA, 02155, USA}
% ...
\affiliation[mm]{Wright Laboratory, Department of Physics, Yale University, New Haven, CT, 06520, USA}

\abstract{Accurate knowledge of electron transport properties is vital to understanding the information provided by liquid argon time projection chambers (LArTPCs). Ionization electron drift-lifetime, local electric field distortions caused by positive ion accumulation, and electron diffusion can all significantly impact the measured signal waveforms. This paper presents a measurement of the effective longitudinal electron diffusion coefficient, $D_L$, in MicroBooNE at the nominal electric field strength of 273.9 V/cm. Historically, this measurement has been made in LArTPC prototype detectors. This represents the first measurement in a large-scale (85 tonne active volume) LArTPC operating in a neutrino beam. This is the largest dataset ever used for this measurement. Using a sample of $\sim$70,000 through-going cosmic ray muon tracks tagged with MicroBooNE's cosmic ray tagger system, we measure $D_L = 3.74^{+0.28}_{-0.29}$ cm$^2$/s.}


%\arxivnumber{1234.56789} % only if you have one

\emailAdd{microboone\_info@fnal.gov}
\date{December 2020}
\maketitle

%%%%%%%%%%% SECTIONS %%%%%%%%%%%%%
\input{sections/01_intro}
\input{sections/02_method}
\input{sections/03_measurement}
\input{sections/04_systematics}
\input{sections/07_discussion}

\input{sections/05_conclusions}

\input{sections/06_acknowledgements}

%%%%%%%%%%%% BIBLIOGRAPHY %%%%%%%%%%%%%%
\newpage
\bibliographystyle{JHEP}
\bibliography{main}

\newpage

%%%%%%%%%%% APPENDICES %%%%%%%%%%%%%%%%
\appendix
\input{appendices/a01_diffusionplotdetails}
%\newpage
\input{appendices/ao2_t0taggingpotential}
%\newpage

\end{document}

%% file: sections/01_intro.tex
\section{Introduction}
\label{sec:intro}

\begin{figure}[!ht]
    \centering
    \includegraphics[width=0.95\textwidth]{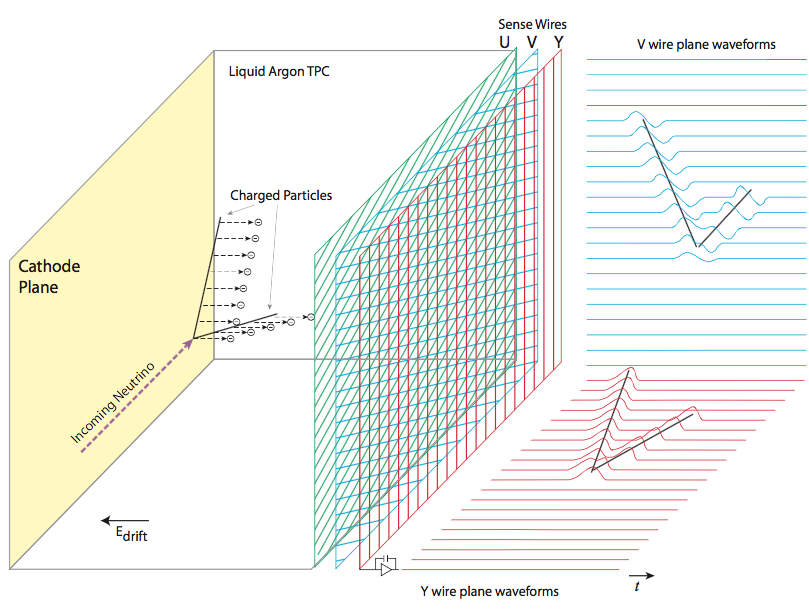}
    \caption{Illustration of the operating principle of a LArTPC. Interactions in the TPC produce charged particles which ionize the argon atoms. These ionization electrons then drift to the anode under the influence of an electric field. Signals are collected from three sense wire planes which act as the system anode. The MicroBooNE coordinate system is such that the $x$-direction is in the direction of the electric field, the $y$-direction is along the vertical axis, and the $z$-direction is in the direction of the neutrino beam. Image credit: reference \cite{microboone2017design_construction}.}
    \label{fig:uboone_cartoon}
\end{figure}

Accurate knowledge of electron transport properties in liquid argon is vital to understanding collected signals in liquid argon time projection chambers (LArTPCs). In LArTPCs, charged particles traversing the detector volume liberate a cloud of ionization electrons from the argon atoms that then drift toward the anode readout plane under the influence of an applied electric field (figure \ref{fig:uboone_cartoon}). During electron transport, multiple processes modify the electron cloud. Slow-drifting Ar$^+$ ions cause distortions in the local electric field (E-field), a process termed the Space Charge Effect (SCE) \cite{abratenko2020measurement}. Ionization electrons may recombine with argon atoms (``recombination''), a process which depends on both the local density of ionization electrons and the local E-field strength and results in an attenuated signal. Ionization electrons can also attach to electronegative contaminants such as O$_2$ and H$_2$O, attenuating the collected signal as a function of drift time. Finally, electron diffusion acts to spread the ionization clouds as a function of drift time. 

%\begin{figure}[b!]
%    \centering
%    \includegraphics[width=1.0\textwidth]{img/intro/ub_angles.pdf}
%    \caption{Diagram of the MicroBooNE coordinate system and wire planes. The beam travels along the $z$-direction, while ionization electrons drift in the decreasing $x$-direction. $y$ denotes the vertical direction. The angles $\theta_{xz}$ and $\theta_{yz}$ denote the angle of a reconstructed object (i.e., track or shower) with respect to the beam direction in the $xz$- and $yz$-planes respectively. The top half shows a side view of the TPC through the anode plane, while the bottom half shows a top-down view. The colored lines (dots) on the top (bottom) half represent the three readout wire planes.}
%    \label{fig:ub_coordinates}
%\end{figure}

\begin{figure}
    \centering
    \includegraphics[width=0.6\textwidth]{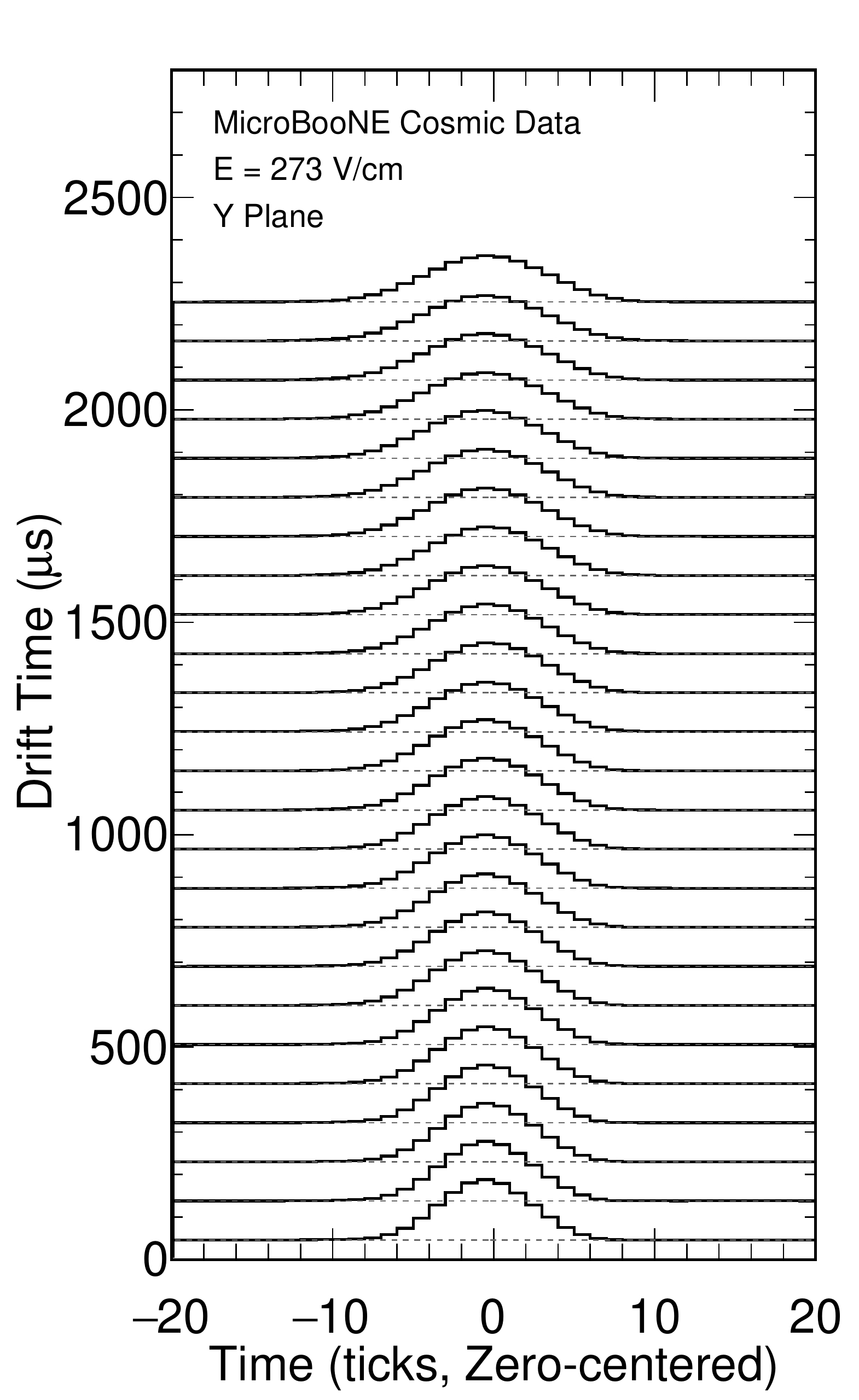}
    \caption{Visualization of the impact of $D_L$ on signal waveforms as a function of drift time. The waveform peak times have been shifted in order to align with one another. One time tick is equivalent to 0.5~$\mu$s. Each waveform displays the deconvolved ADC count, arbitrarily scaled.}
    \label{fig:diffusionDemo}
\end{figure}

Electron diffusion is non-isotropic under the influence of an electric field \cite{Atrazhev:1998,Cennini:1995ha,li2016measurement} and is split into components which are transverse and longitudinal to the E-field. The transverse component, $D_T$, impacts the spatial resolution of a given LArTPC in the plane parallel to the readout wire plane (the $yz$-plane in the MicroBooNE coordinate system, shown in the top half of figure \ref{fig:ub_diagram}). Similarly, the longitudinal component, $D_L$, impacts the spatial resolution along the drift coordinate (perpendicular to the wire plane as shown in the bottom half of figure \ref{fig:ub_diagram}) broadening the signal waveforms as a function of drift time as shown visually in figure \ref{fig:diffusionDemo}. For particles near the anode, where the drift time is low, the signal waveform is relatively tall and narrow. As the drift time increases, the pulses become shorter and broader. 

Few measurements of $D_L$ currently exist in liquid argon. In 1994, the ICARUS collaboration reported measurements of $D_L$  at E-fields ranging from 100 to 350 V/cm using a three-ton LArTPC with a maximum drift distance of 42 cm \cite{Cennini:1995ha}. A more recent but preliminary measurement using the ICARUS T600 detector is reported in reference \cite{Torti:2017}. Li et al. from Brookhaven National Lab (BNL) reported measurements between 100 and 2000 V/cm in 2015 using a laser-pulsed gold photocathode with drift distances ranging from 5 to 60 mm \cite{li2016measurement}. The ICARUS results show good agreement with the prediction of Atrazhev and Timoshkin \cite{Atrazhev:1998}, while the results of Li et al. are systematically higher than both. Figure \ref{fig:dl_summary} summarizes the current published world data for $D_L$ measurements. 

%\begin{figure}
%    \centering
%    \includegraphics[width=0.8\textwidth]{img/diffusionplotdetails/dl_summary_noub.pdf}
%    \caption{Summary of world data for longitudinal electron diffusion in liquid argon. The orange-dashed curve shows the theory prediction \cite{Atrazhev:1998}, the blue dot-dashed curve shows the parametrization from Li et al. \cite{li2016measurement}, and the red and dark blue points show the ICARUS \cite{Cennini:1995ha} and Li et al. measurements, respectively. Details of this plot can be found in Appendix \ref{app:diffusion_plot_details}. Note that the ICARUS error bars ($\pm$ 0.2 cm$^2$/s) are covered by the data point.}
%    \label{fig:dl_summary_no_uboone}
%\end{figure}

\section{The MicroBooNE Detector}

The MicroBooNE detector \cite{microboone2017design_construction} is a LArTPC currently running as part of the short-baseline neutrino (SBN) program at Fermi National Accelerator Laboratory. It is exposed to neutrinos from the Booster Neutrino Beam and is situated on-axis 463 m downstream of the neutrino production target. The MicroBooNE experiment is the longest running large-scale LArTPC in the world to date, having first started data taking in August 2015. The detector provides an excellent opportunity to develop techniques for performing detector physics measurements in a running LArTPC that will benefit future large-scale LArTPCs such as the upcoming Deep Underground Neutrino Experiment (DUNE) \cite{Abi:2020wmh}. The MicroBooNE TPC measures 10.37~m along the beam direction, 2.56~m in the drift direction, 2.33~m in the vertical direction, and contains 85 tonnes of liquid argon in the active volume. The MicroBooNE TPC operates at an E-field of 273.9~V/cm and the liquid argon temperature is held stably at $89.4 \pm 0.2$~K. Under nominal operating conditions, the maximum ionization electron drift time is 2.3 ms. The anode wire plane consists of 8256 wires separated into two induction planes (U and V planes) each with 2400 wires angled at $\pm60^{\circ}$ to the vertical, and a collection plane (Y plane) with 3456 wires and oriented vertically. Collected signals are sampled at a frequency of 2 MHz; one time sample (``tick'') is 0.5 $\mu$s.

\begin{figure}[!hb]
    \centering
    \includegraphics[width=0.95\textwidth]{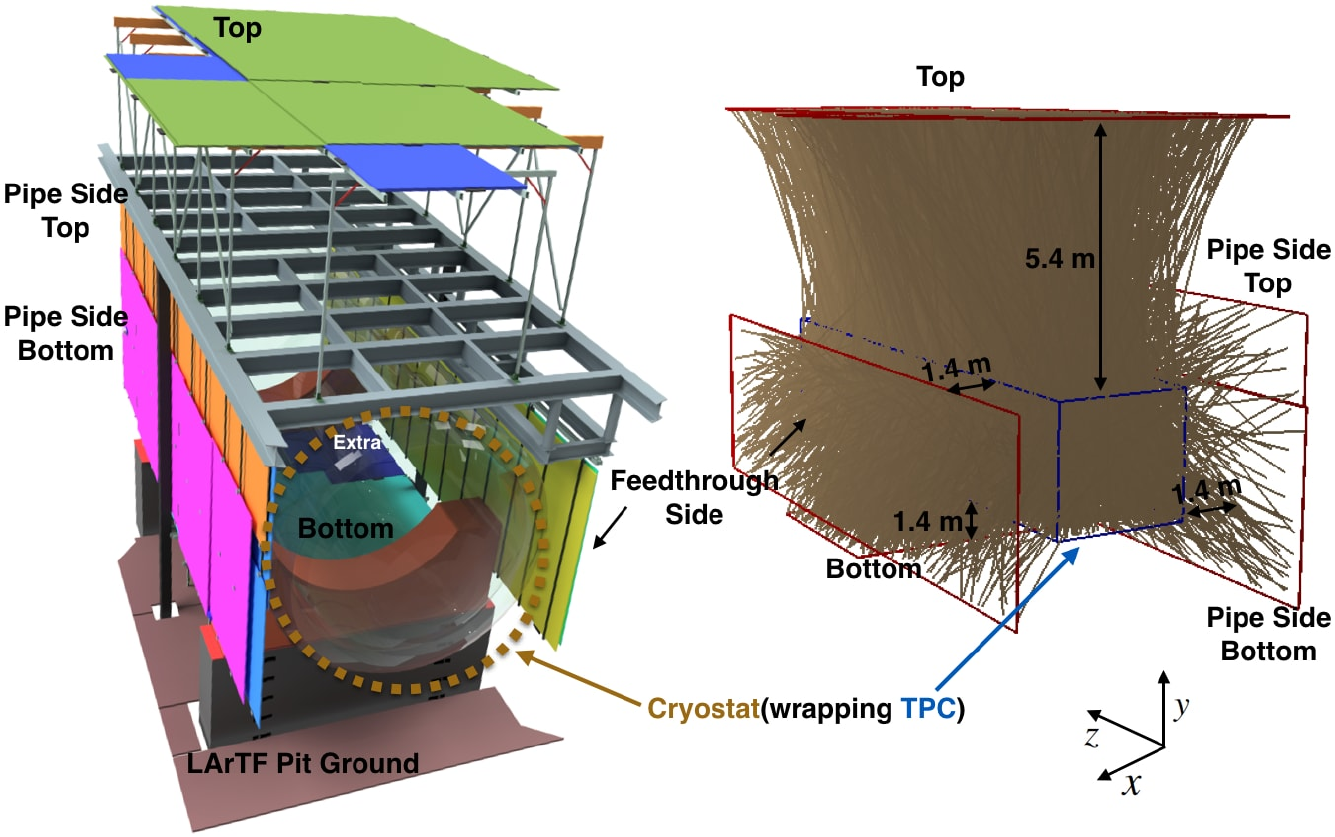}
    \caption{Positioning of the CRT planes in MicroBooNE, taken from reference \cite{crt2019design}. Left: placement of CRT modules with respect to the TPC. Right: simulation of cosmic muons tagged with the CRT system. }
    \label{fig:crt_design_cartoon}
\end{figure}

The MicroBooNE detector was upgraded with a Cosmic Ray Tagger (CRT) system \cite{crt2019design} which was installed in October 2017. The position of the CRT planes with respect to the TPC is shown in figure \ref{fig:crt_design_cartoon} along with a simulation of CRT-tagged cosmic muon tracks. Due to space constraints in the Liquid Argon Test Facility that houses the MicroBooNE detector, there are no CRT planes at the upstream or downstream ends of the detector. Additionally, the top plane is 5.4 m from the top face of the detector in order to accommodate detector electronics racks. As built, the CRT provides a maximum solid angle coverage of 85\%. 

This work describes the measurement of longitudinal ionization electron diffusion in the MicroBooNE detector utilizing a novel technique using cosmic-ray muons tagged by the CRT system. The data used in this analysis uses MicroBooNE cosmic-ray data taken between October 27, 2017 and March 13, 2018. 

%% file: sections/02_method.tex
\section{Method}\label{sec:method}

To first order, the relationship between the time-width of a signal pulse at a given time $t$, $\sigma_t(t)$, and $D_L$ can be parametrized \cite{Cennini:1995ha, li2016measurement} as

\begin{equation}
    \sigma_t^{2}(t)\simeq \sigma_t^2(0) + \left(\frac{2D_L}{v_d^{2}}\right)t,
    \label{eqn:diffusionEqn}
\end{equation}
where $v_d$ is the drift velocity and $\sigma_t^2(0)$ is added to account for the Gaussian noise filter used during waveform deconvolution \cite{signal_processing_paper1} which enforces a minimum width for the pulses. The expected minimum width is $\sigma_t^2(0)\sim 1.96 \ \mu s^2$. Because equation (\ref{eqn:diffusionEqn}) is an approximation, $D_L$ is actually an \emph{effective} diffusion coefficient that contains a small contribution from transverse diffusion (see section \ref{sec:systematics}). Equation (\ref{eqn:diffusionEqn}) assumes a constant $v_d$. However, due to the abundant cosmic ray flux in MicroBooNE caused by its location near the surface, the electric field varies as a function of position in the detector due to SCE. This means that $v_d$ also changes throughout the detector volume. MicroBooNE has measured the values of $v_d$ as a function of TPC position using electric field maps determined using UV laser data \cite{chen2019_drift_velocity}. Because equation (\ref{eqn:diffusionEqn}) captures the size of the electron cloud at the point of measurement, it is important to use the value of the drift velocity at the location of that measurement. Specifically, the signal processing removes the electronics response and the field shaping and returns a measured time distribution that corresponds to the  arrival time of the electrons at $x=0$ (the first induction plane) convoluted with a Gaussian low-pass filter function that removes high frequency noise. Thus, the mean drift velocity at $x=0$, $v_d = 1.076$ mm/$\mu$s, is used for the measurement of $D_L$ in the MicroBooNE data. When measuring $D_L$ from our simulation samples (see section \ref{subsec:val_particle_gun}), we use the nominal simulated $v_d$ value of 1.098 mm/$\mu$s, since the simulated signal deconvolution assumes the ionization electrons drift at this velocity across the volume. Systematic uncertainties associated with the drift velocity are considered in section \ref{subsec:syst_vd_sce}.

Although the MicroBooNE E-field varies as a function of position within $273.9^{+12\%}_{-8\%}$ V/cm \cite{abratenko2020measurement, chen2019_drift_velocity} due to space charge effects, equation (\ref{eqn:diffusionEqn}) assumes that the value of $D_L$ is constant and this assumption is built into the analysis. Figure \ref{fig:dl_summary} shows that, within MicroBooNE's E-field variations, the current world data and theoretical expectations for $D_L$ are consistent with an assumption of a constant $D_L$ value in the region of the MicroBooNE E-field. The MicroBooNE nominal $D_L$ simulation value is extracted from the parametrization of Li et al. (blue-dashed curve in figure \ref{fig:dl_summary}) at $E = 273.9$ V/cm and corresponds to a $D_L$ value of 6.40 cm$^2$/s.

\subsection{Event Selection}

The signals collected from the three wire planes undergo processing to remove the detector response\footnote{The measured signal can be considered a convolution of the true signal and the detector response.} \cite{signal_processing_paper1,signal_processing_paper2} resulting in waveforms which are approximately Gaussian. Each waveform is then fitted with a Gaussian function producing a reconstructed hit. High-level reconstructed objects such as tracks, showers, and space points (three-dimensional hits) are created from collections of hits using the Pandora multi-algorithm pattern recognition software \cite{Acciarri:2017hat}.

Due to the linear relationship between the squared-pulse-width in time and ionization electron drift time (equation (\ref{eqn:diffusionEqn})), it suffices to perform a linear fit of $\sigma_t^2$ versus $t$ and extract $D_L$ from the slope. The widths of waveforms (``pulse widths'') are sensitive to more effects than just longitudinal diffusion. Transverse diffusion, the detector response modeling, collinear delta ray production, and the angle of the reconstructed track can all significantly impact the measured time width of the pulse. To minimize the additional broadening from such effects, we place a strict set of requirements on tracks reconstructed from the MicroBooNE data.

\subsubsection{Track Selection}\label{subsec:track_selection}

To measure $D_L$, we use cosmic muons tagged by MicroBooNE's CRT. Using the signals read out from the CRT system along with the start and end points of the reconstructed cosmic track, we can determine the drift time ($t_0$) that a cosmic muon entered the detector. This allows us to use $t_0$ as the track start time to determine the drift time of the waveforms used in the final measurement. Tracks with a known $t_0$ are said to be \emph{$t_0$-tagged.} The CRT itself has an internal precision of $\sim1$~ns \cite{crt2019design}, which translates to a time resolution on the TPC clock of $<1$~$\mu$s. For CRT-tagged tracks with length greater than 50 cm, the $t_0$-tagging efficiency is 56.6\%.  For this analysis, we require that tracks must

\begin{itemize}
    \item have a reconstructed length greater than 50 cm; 
    \item be through-going, meaning that both the start and end points must be within 5 cm of any TPC wall; %after applying SCE spatial corrections;
    \item have $|\theta_{xz}| < 6^\circ$ and $|\theta_{yz}| < 40^\circ$ (figure \ref{fig:ub_diagram}); and
    \item have an average track deflection of less than 6 cm.
\end{itemize}

\begin{figure}
    \centering
    \includegraphics[width=\textwidth]{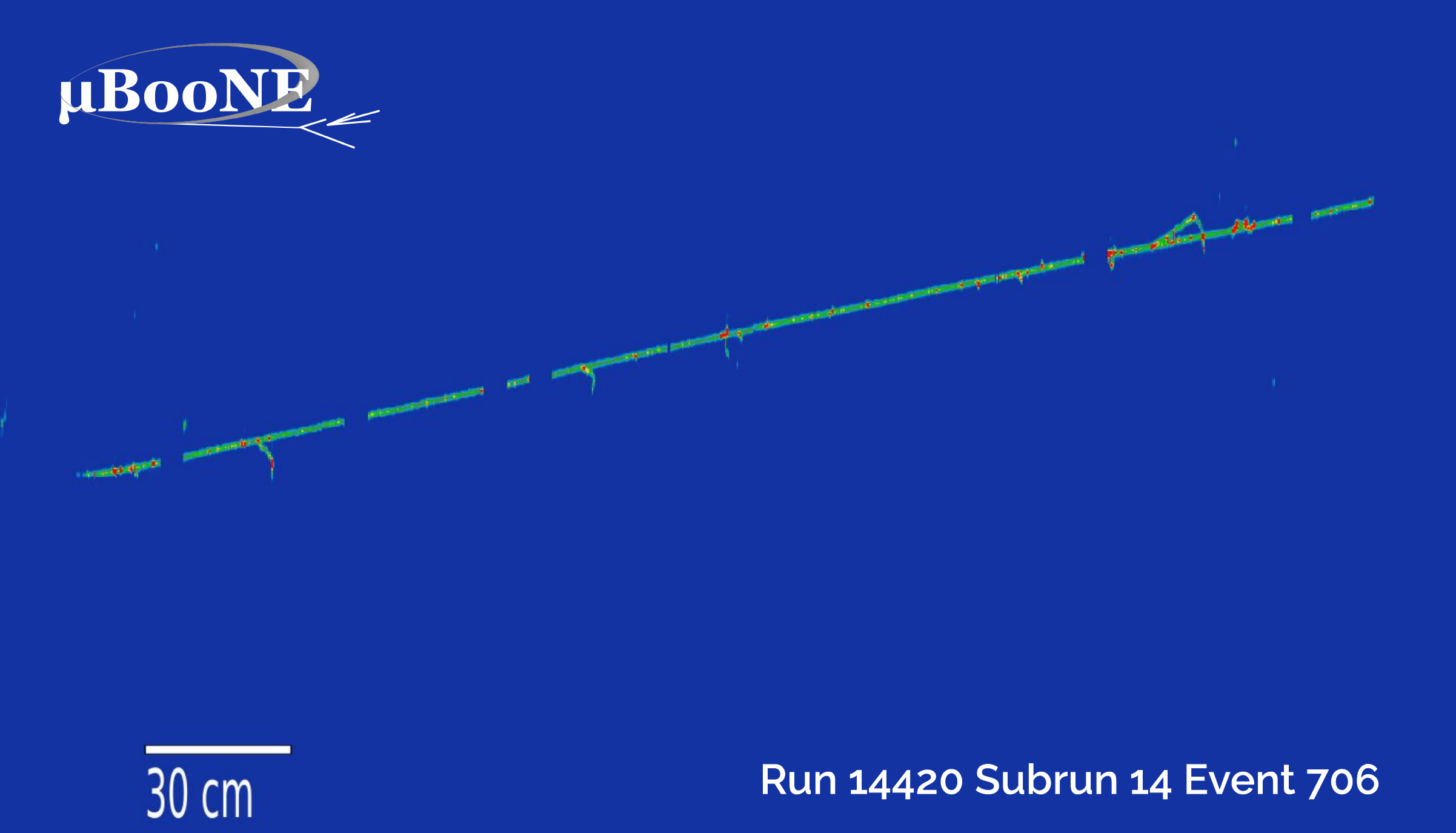}
    \caption{Event display of a track that passes the diffusion track selection outlined in section \ref{subsec:track_selection}. This image shows deconvolved data from the MicroBooNE collection plane. The wire number increases in the horizontal direction from left to right, while the vertical direction shows the time at which each charge deposition was collected, with charge near the top being collected later than at the bottom.}
    \label{fig:evd_filter}
\end{figure}

The track length requirement ensures track reconstruction quality and reduces potential track mis-identification of shorter tracks or shower-like objects. We require through-going tracks as an additional reconstruction quality check. The strict angular selection is designed to mitigate additional pulse width broadening due to the combined effects of track angle, $D_T$, and the detector response modeling (see figure 9 of reference \cite{signal_processing_paper1}) particularly in the $xz$-plane. As $\theta_{xz}$ increases, so does the intrinsic spread in $x$ of the ionization position distribution. A stringent $\theta_{xz}$ requirement therefore mitigates this effect while providing a sufficient number of waveforms to perform the analysis. $\theta_{yz}$, on the other hand, impacts pulse height rather than pulse width, so we choose a looser requirement for that angle. Finally, as a measure of track straightness, we use the average deflection defined as the average transverse distance between each point along the track and a straight line connecting the track start and end points. Track angles are determined using the track starting direction, but, in some cases, the track can significantly deviate from this starting direction due to both the SCE and multiple Coulomb scattering \cite{Abratenko:2017nki}. This requirement therefore ensures that tracks remain relatively forward-going. An event display of a selected track is shown in figure \ref{fig:evd_filter}.

\begin{figure}
    \centering
    \includegraphics[width=0.9\textwidth]{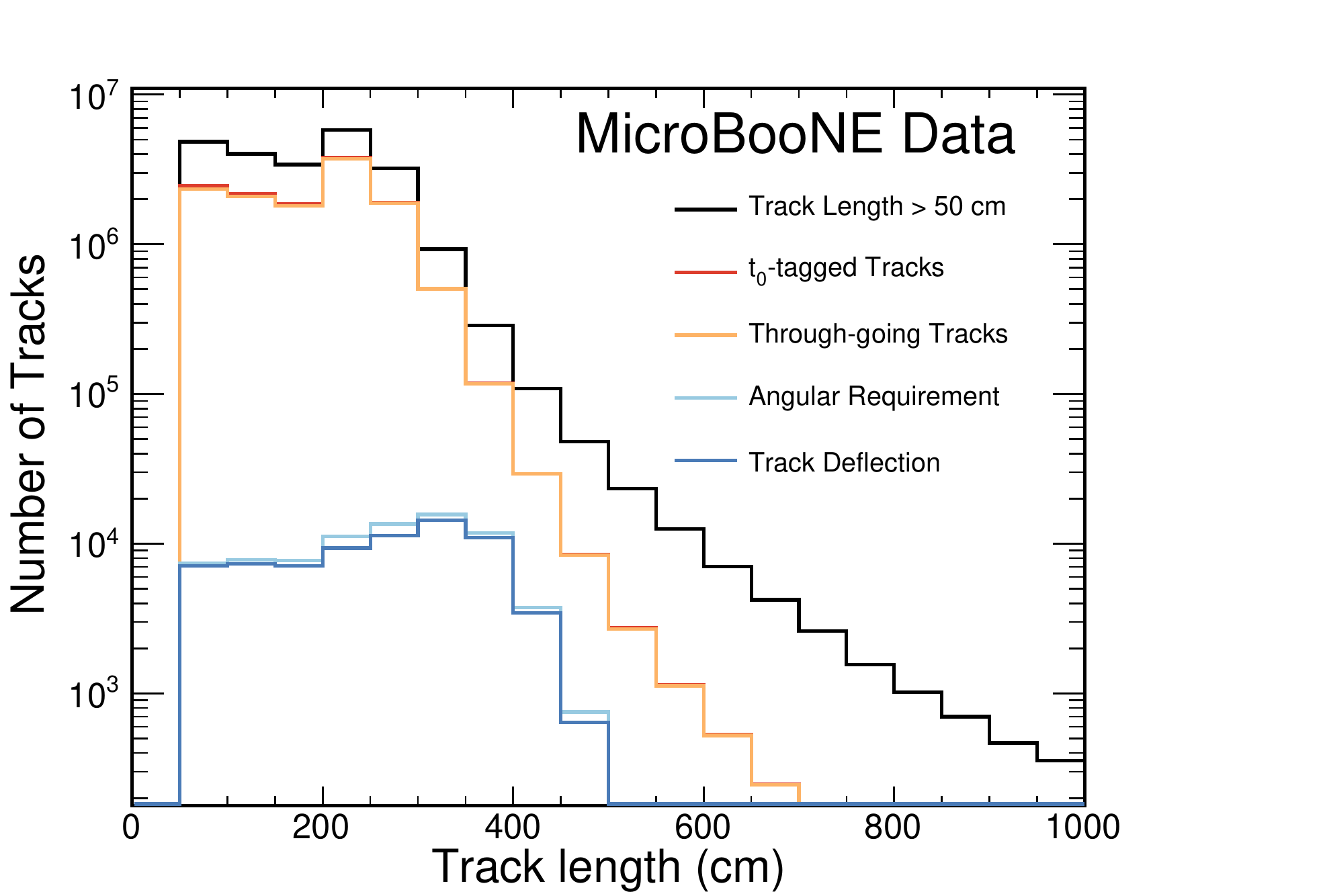}
    \caption{Track length distributions at each stage of the track selection. The peak around 230 cm in the orange and black curves corresponds to the height of the TPC since most CRT tracks traverse the detector top-to-bottom.}
    \label{fig:track_lengths}
\end{figure}

Track length distributions at each stage of the selection are shown in figure \ref{fig:track_lengths}, while the selection efficiencies and number of selected tracks are shown in table \ref{tab:selection_cut_eff}. The requirements on the track angle are the least efficient, reducing the number of selected tracks by two orders of magnitude. The final selection contains $\sim$70,000 tracks and each track can have hundreds of waveforms. This provides an ample number of waveforms to perform the analysis.

\begin{table}[ht]
    \centering
    \begin{tabular}{l|c|c|c}
         \textbf{Selection Requirement} & \textbf{No. Tracks} & \textbf{Relative Efficiency} & \textbf{Absolute Efficiency}\\
         \hline 
         Reconstructed tracks & 5.27$\times10^7$ & 100\% & 100\%\\
         Track length $>$ 50 cm & 2.27$\times10^7$ & 43.1\% & 43.1\%\\
         Track is $t_0$-tagged & 1.28$\times10^7$ & 56.4\% & 24.3\% \\
         Track is through-going& 1.25$\times10^7$ & 97.7\% & 23.7\% \\
         Track meets angular requirement & 79,896 & 0.64\% & 0.15\% \\
         Track meets deflection requirement & 71,698 & 89.7\% & 0.14\%\\
    \end{tabular}
    \caption{Selection efficiencies after each selection requirement and number of selected tracks. Relative efficiencies are calculated relative to the number of tracks at the previous stage of the selection.}
    \label{tab:selection_cut_eff}
\end{table}

\subsubsection{Waveform Selection}\label{subsec:waveform_selection}

The pulse widths in this analysis are extracted from deconvolved waveforms, low-level data products which attempt to recover a ``true'' signal by deconvolving the raw signal measured at the anode wires with the detector response. The MicroBooNE detector response is modeled as a convolution of a \emph{field} response and an \emph{electronics} response. The field response describes the charge induced on one anode-plane wire by a single ionization electron, while the electronics response describes the impact on the signal waveform due to shaping and amplification during signal readout \cite{signal_processing_paper2, Acciarri:2017sde}. The deconvolution process also applies a Gaussian low-pass noise filter to mitigate the effects of electronics noise \cite{Acciarri:2017sde}. In the case of a simple one-dimensional deconvolution, the true deconvolved frequency-space signal, $S(\omega)$, can then be modeled as 

\begin{equation}\label{eqn:deconvolution_response}
    S(\omega) = \frac{M(\omega)}{R(\omega)}F(\omega),
\end{equation}
where $M$ is the measured signal, $R$ is the detector response, and $F$ is the Gaussian noise filter \cite{signal_processing_paper1}. The MicroBooNE deconvolution is two-dimensional; it is applied in both time and wire space. The details can be found in references \cite{signal_processing_paper1,signal_processing_paper2}. The result is a signal waveform with a distinct region of interest preserved around signal peaks that exceed a predefined threshold value.\footnote{For the U, V, and Y planes, these threshold values are 350, 500, and 300 electrons/tick, respectively \cite{signal_processing_paper1}.} 

As with the reconstructed tracks, we place a set of requirements on the reconstructed hits to ensure waveform quality. While the final $D_L$ measurement uses deconvolved waveforms rather than reconstructed hits, hit information is easily accessible and can be used as a proxy for the shape of the underlying waveform. We require reconstructed waveforms for which the reconstructed hits

\begin{figure}
    \centering
    \includegraphics[width=1.0\textwidth]{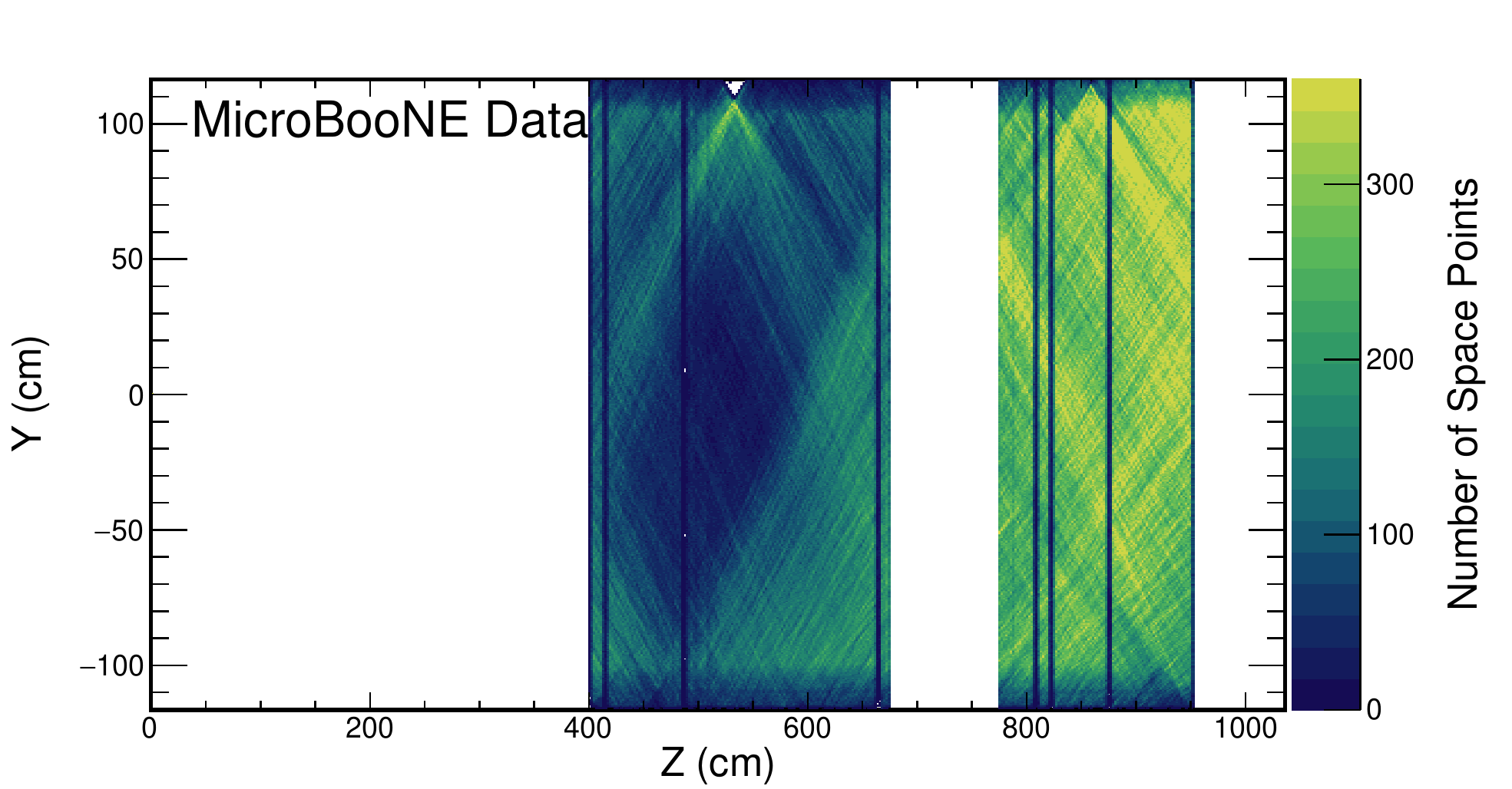}
    \caption{$yz$-position distribution of reconstructed space points corresponding to selected hits on the collection plane. The majority of the selected hits are in the downstream portion of the detector due to geometric effects along with the track selection. The empty region at the top of the detector around $z\sim550$ cm is due to the overlap of two dead regions on the two induction planes. 3D reconstructed objects such as space points require charge to have been measured on at least two wire planes.}
    \label{fig:spacepoint_yz}
\end{figure}

\begin{itemize}
    \item have been fit to a single Gaussian distribution;
    \item satisfy a goodness-of-fit (GoF) requirement that was tuned to be highly efficient at keeping Gaussian-like signals, while removing waveforms which are distorted due to overlapping hits (primarily due to delta rays); and
    \item have a $z$-position between 400 cm $< z <$ 675 cm or 775 $< z <$ 951 cm.
\end{itemize}
Requiring the waveform to have been fit to a single Gaussian distribution removes hits that are contaminated with other charge depositions, particularly those due to delta ray production along the reconstructed track. The hit GoF test ensures that the waveform shape is reasonably Gaussian; we model electron diffusion as a Gaussian process, and the deconvolution uses a Gaussian noise filter. We expect the waveforms to follow this shape as well. Finally, we apply a hit fiducial volume along the $z$-direction. The first induction plane in MicroBooNE is known to have a region of shorted wires in the upstream half of the TPC \cite{signal_processing_paper1, Acciarri:2017sde}; requiring hit positions to be at least 400 cm from the upstream end of the TPC removes this region from consideration. The downstream portion of the detector volume is impacted by SCE \cite{chen2019_drift_velocity}, and we remove that region as well. Finally, we ignore the region between 675 and 775 cm in $z$ to avoid a region of dead wires in the collection plane. Figure \ref{fig:spacepoint_yz} shows the $yz$-position distribution of reconstructed space points corresponding to the selected hits. The waveform fiducial volume removes slightly more than half of the TPC volume with most of the selected waveforms coming from $z > 800$ cm due to the detector geometry combined with the requirement that reconstructed tracks have a shallow $\theta_{yz}$ and be through-going.

\begin{figure}
    \centering
    \begin{subfigure}{0.49\textwidth}
        \includegraphics[width=\textwidth]{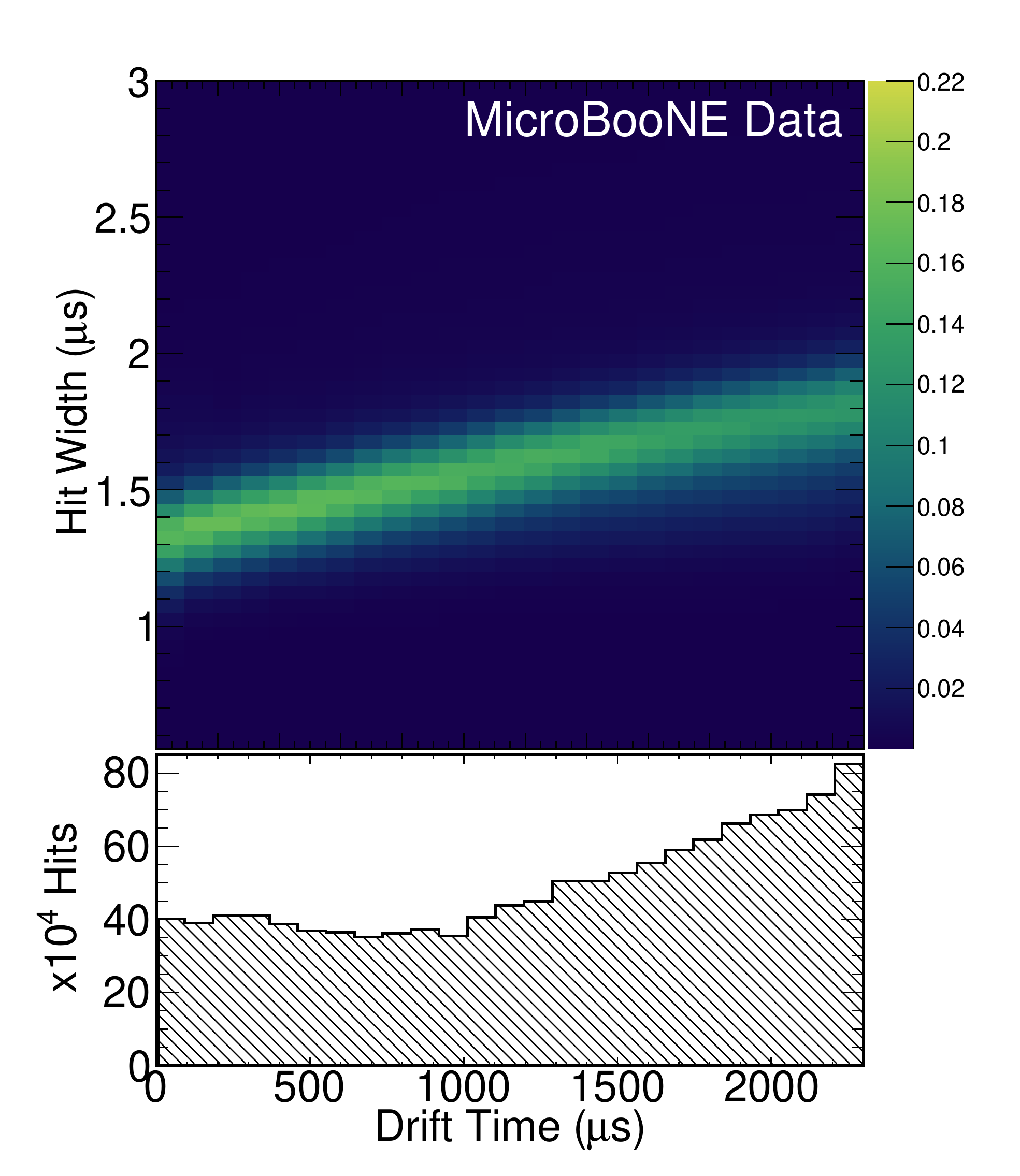}
        \caption{No requirement}
        \label{subfig:dynamic_sigma_precut}
    \end{subfigure}
    \begin{subfigure}{0.49\textwidth}
        \includegraphics[width=\textwidth]{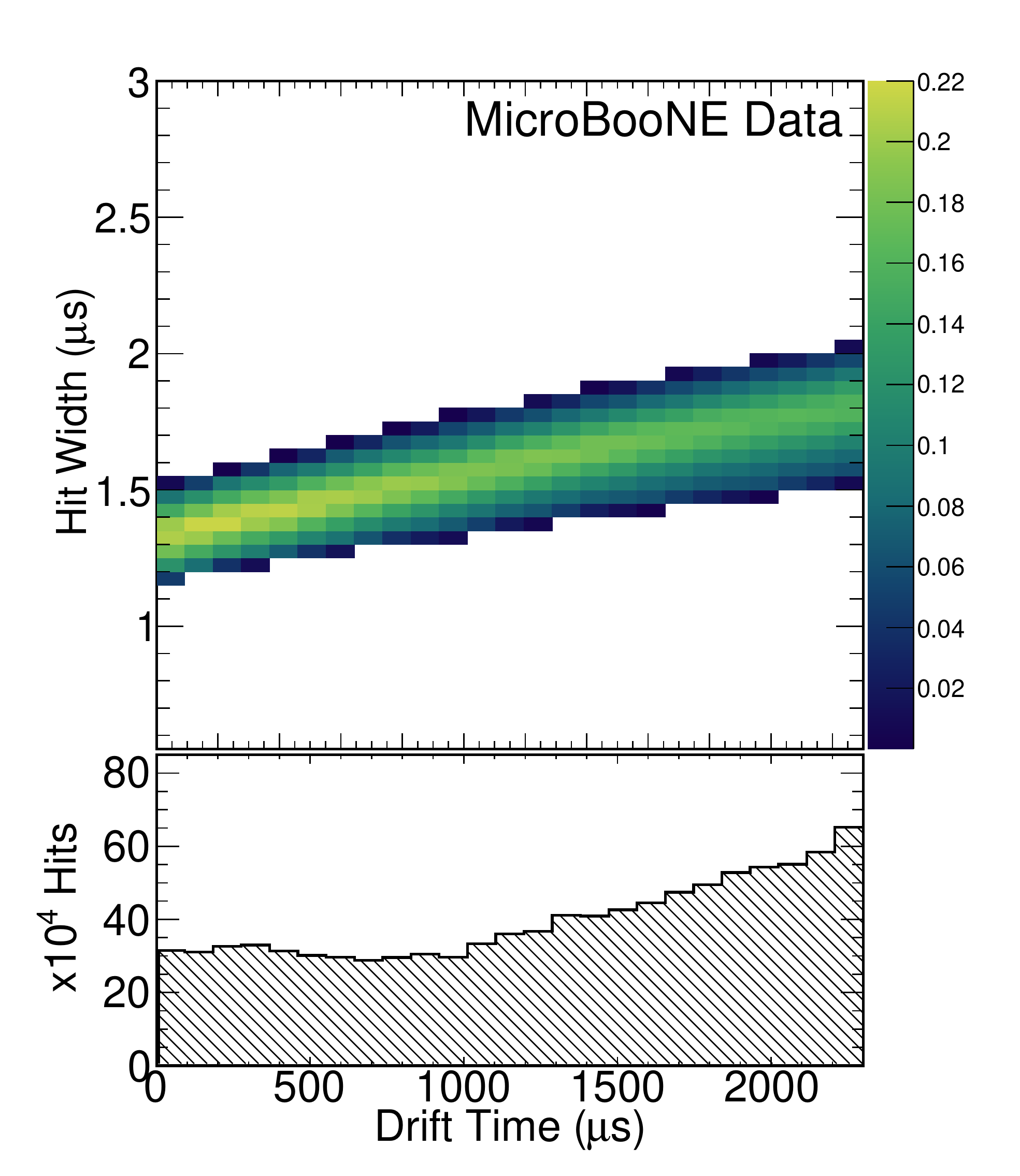}
        \caption{1$\sigma$ requirement}
        \label{subfig:dyanmic_sigma_postcut}
    \end{subfigure}
    \caption{(a) Distribution of hit widths vs. drift time. (b) The same distribution after requiring hits be within one standard deviation of the median value in each drift bin as described in the text. Each bin of drift time has been area normalized in the two dimensional histogram so that the structure is more visible. The bottom histograms show the number of hits collected in each bin of drift time.}
    \label{fig:dynamic_sigma_cut}
\end{figure}

In addition to the criteria listed above, we place an additional requirement that the hit width of each individual waveform be representative of the pulse width distribution in its corresponding bin of drift time. To do so, we reject all waveforms whose hit widths fall outside of a one standard deviation region around the median value in that drift bin as shown in figure \ref{fig:dynamic_sigma_cut}. The dark blue regions in figure \ref{subfig:dynamic_sigma_precut} show that many hit widths differ significantly from the median value in that drift bin largely due to effects such as unresolved delta rays, misreconstruction, and the statistical nature of diffusion. Adding this requirement reduces the bias in this measurement. To investigate whether our specific choices of using the median and standard deviation of the distribution introduce any bias to the measurement, we investigated using the peak of the distribution rather than the median, and placing different requirements on how close a waveform must be to the median value in the drift bin. Changes to the measured $D_L$ value have been found to be at the sub-percent level. 

%This leads to long ``tails'' in the distribution of $\sigma_t^2$ in each bin which could bias the width of the resultant summed waveforms in each bin. 

\subsection{Extraction of \texorpdfstring{$D_L$}{DL}}\label{subsec:extraction_DL}

\begin{figure}
    \centering
    \includegraphics[width=0.32\textwidth]{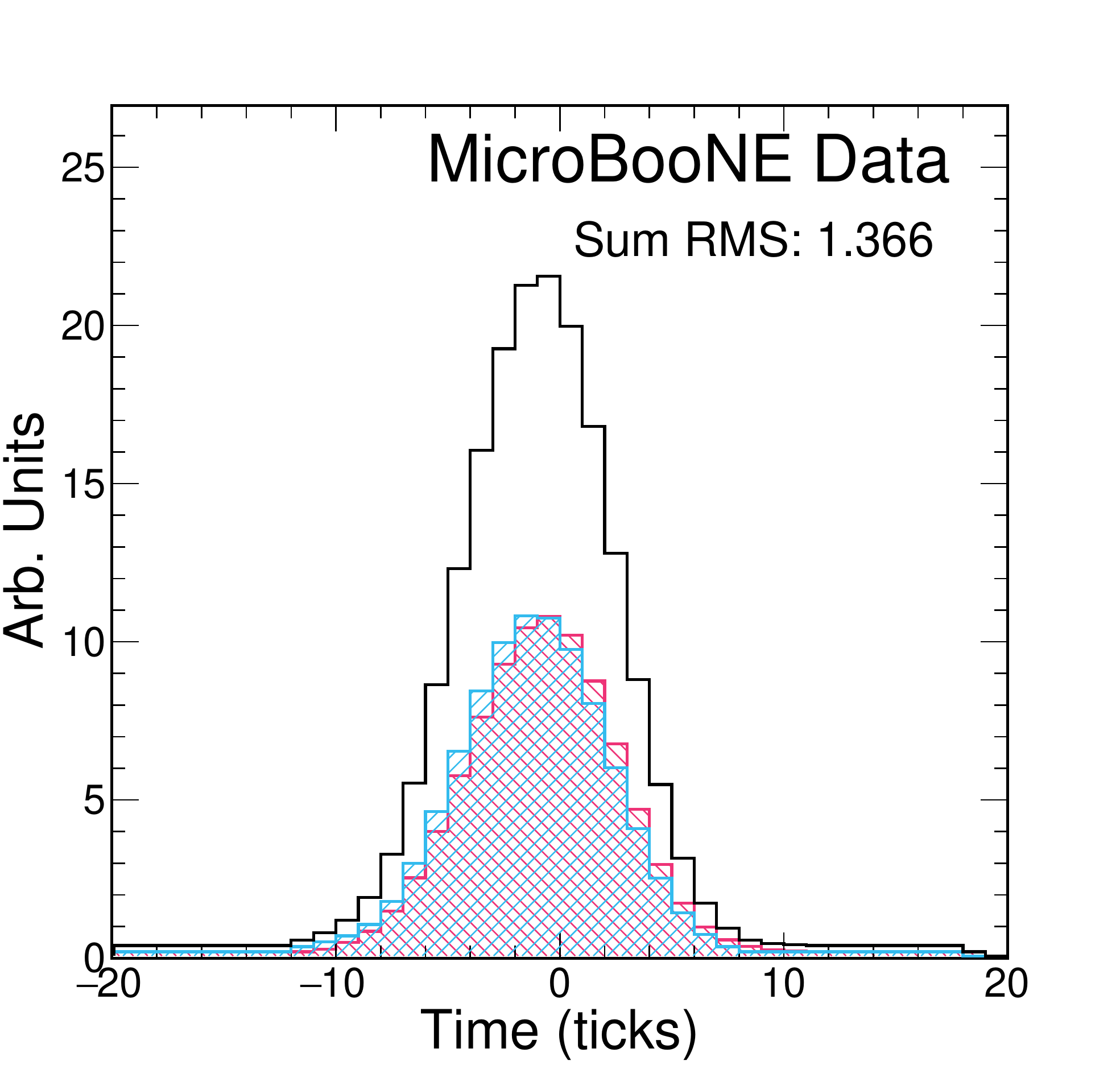}
    \includegraphics[width=0.32\textwidth]{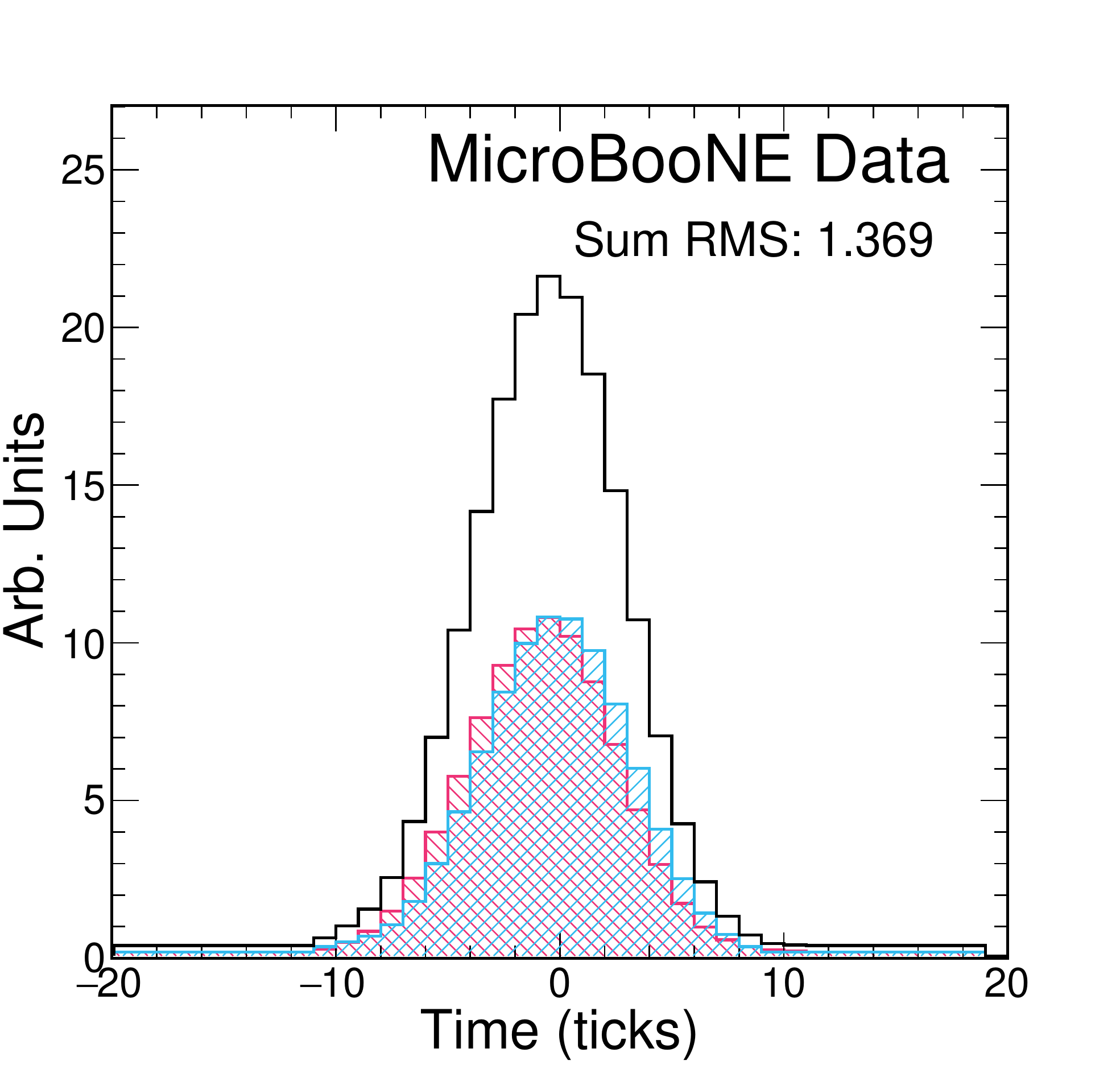}
    \includegraphics[width=0.32\textwidth]{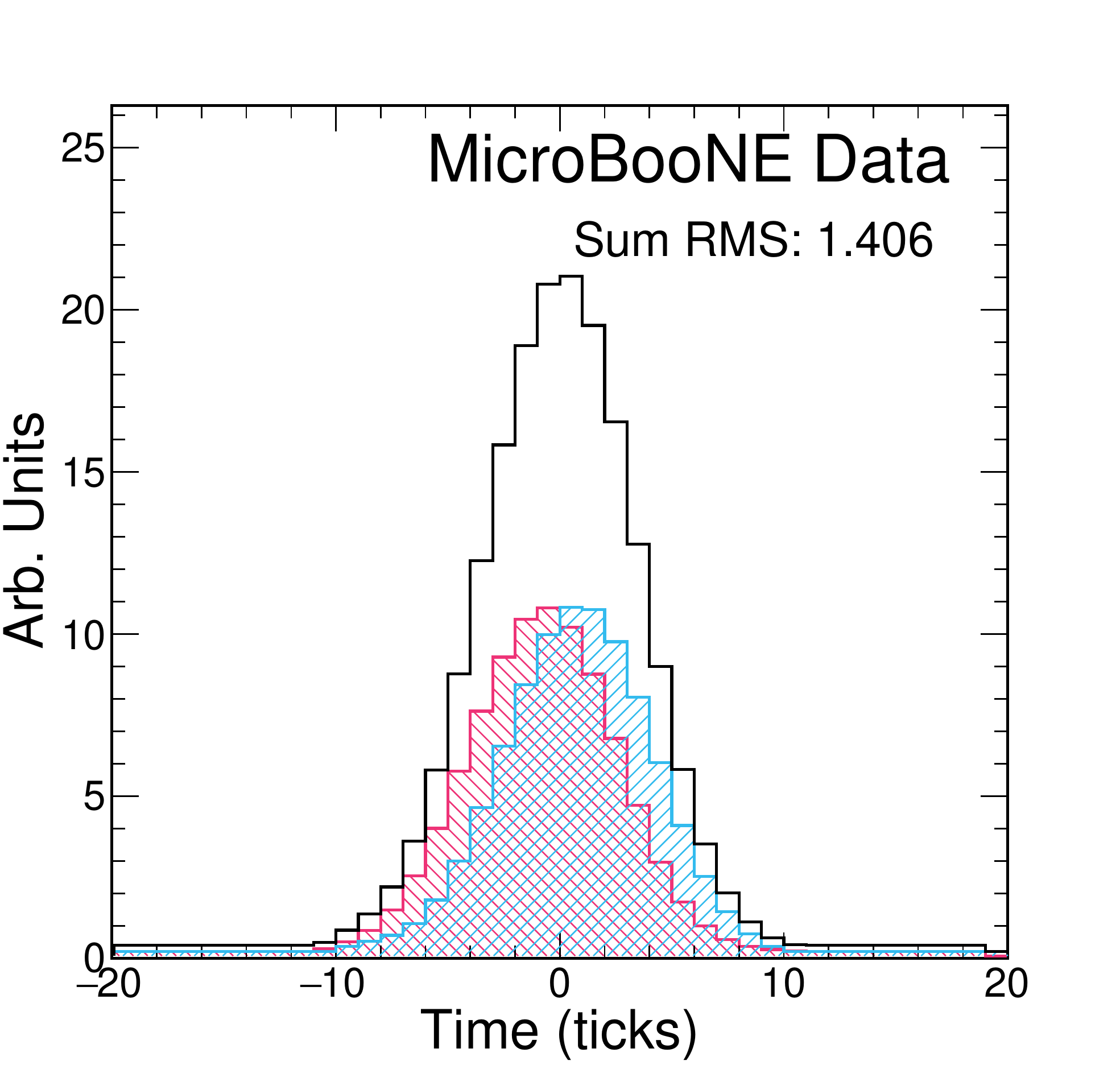}
    \caption{Illustration of the waveform summation technique employed in this analysis. The cyan waveform is iteratively shifted from -5 to +5 ticks in increments of one tick. At each iteration, the cyan waveform is added to the magenta waveform and the RMS of the summed waveform (black) is calculated. In this simplified example, the cyan waveform is shown shifted by -1, 0, and +1 ticks. In this case, the left-hand configuration would be selected.}
    \label{fig:waveform_averaging_summation} 
\end{figure}

\begin{figure}
    \centering
    \includegraphics[width=0.8\textwidth]{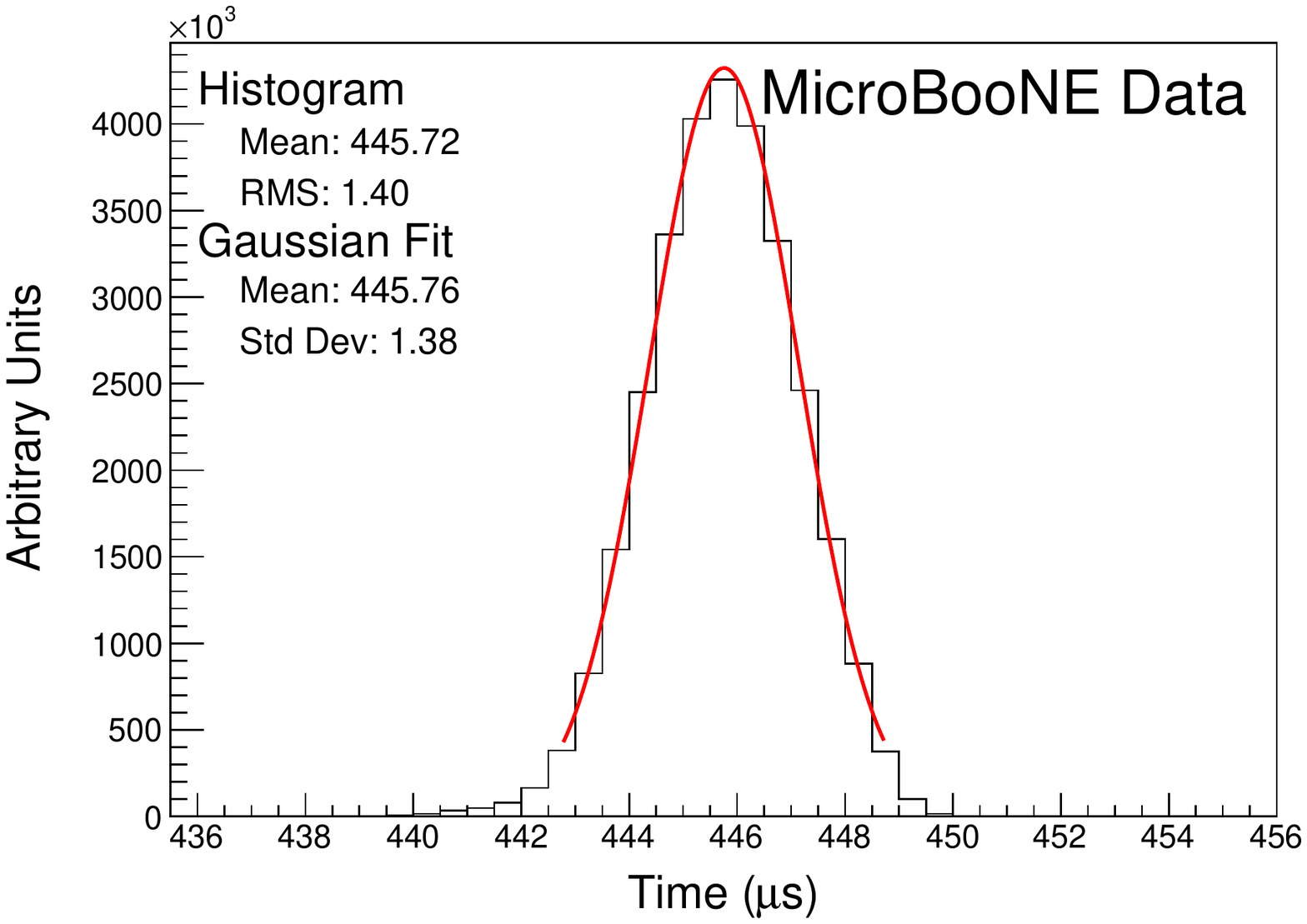}
    \caption{Sample summed waveform with Gaussian fit. $\sigma_t$ is extracted from the standard deviation of the fit and $t$ from the mean. This waveform is taken from the first drift bin on the collection plane.}
    \label{fig:summed_waveform_with_fit}
\end{figure}

The electron drift time in MicroBooNE ranges from 0 to 2300 $\mu$s, which we split into 25 bins. At the nominal drift velocity of 1.098 mm/$\mu$s, each bin corresponds to roughly 10 cm of drift distance. Within each of the 25 drift time bins, we employ a waveform summation technique to obtain a single representative waveform of that bin. To account for time offsets between waveforms we iteratively shift each additional waveform from -5 ticks to +5 ticks relative to the center of the summed waveform and choose the configuration which minimizes the RMS of the resultant summed waveform. An example of this process is shown in figure \ref{fig:waveform_averaging_summation}, and a sample summed waveform is shown in figure \ref{fig:summed_waveform_with_fit}. The summed waveform retains a Gaussian shape, without a significant additional broadening due to the waveform summation method; see section \ref{sec:syst_waveform_summation}.

Once we have a summed waveform in each bin, we fit a Gaussian to that summed waveform, taking the standard deviation as our measure of $\sigma_t$, and the mean as $t$. We then plot $\sigma_t^2$ vs $t$, and extract $D_L$ from the slope of this fit. Figure \ref{fig:summed_waveform_with_fit} shows a sample summed waveform with the Gaussian fit drawn on top. It is clear that the underlying distribution is not perfectly Gaussian, but when restricted to the region around the peak of the distribution, the Gaussian functional form is a good estimator of the width of the distribution. The statistical uncertainty on $\sigma_t$ is negligible due to the large number of waveforms used in each drift bin.

%% file: sections/03_measurement.tex
\section{Measurement of Longitudinal Electron Diffusion}\label{sec:measurement}

\subsection{Method Validation on Simulated Samples}\label{subsec:val_particle_gun}

\begin{figure}
    \centering
    \includegraphics[width=\linewidth]{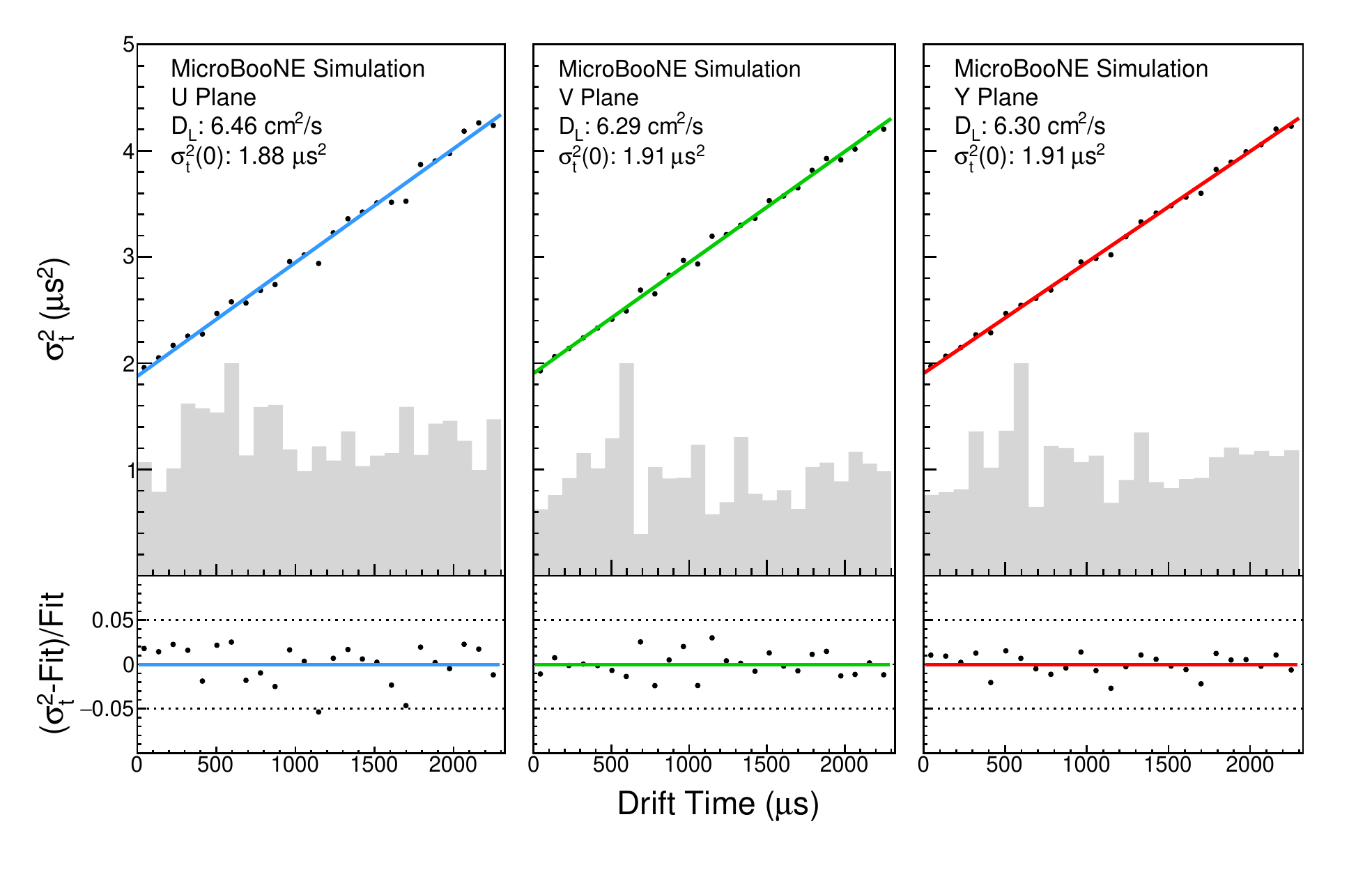}
    \caption{Plots of $\sigma_t^2$ versus $t$ for simulated muons generated within the angular selection values of $\theta_{xz} = \pm6^{\circ}$ and $\theta_{yz} = \pm40^{\circ}$. The shaded histograms show the area-normalized distributions of the number of waveforms in each bin. The bottom plots show the fit residuals of each point. The induction planes are used only to estimate systematic uncertainties (see section \ref{sec:systematics}).}
    \label{fig:d3planediffresultSim}
\end{figure}

To validate the method described in section \ref{sec:method}, we use simulated samples containing only a single muon. These simplified samples contain 500 events, each with exactly one muon track and no backgrounds. The muon tracks are generated precisely in-time with the beam, so there is no potential bias from $t_0$ mis-tagging. They populate the detector volume uniformly and have a fixed momentum of 1 GeV$/c$, with an angular coverage of $\theta_{XZ} = \pm6^{\circ}$. Figure \ref{fig:d3planediffresultSim} shows the resultant plots of $\sigma_t^2$ vs.\ $t$ on each wire plane for simulated single muons within the angular selection values listed in section \ref{subsec:track_selection}. For each plane, the top plot shows the linear fit and an area-normalized histogram of the number of waveforms in each bin; the bottom plot shows the fit residuals. As discussed in the previous section, each point on the plots in figure \ref{fig:d3planediffresultSim} represents the standard deviation of a Gaussian fit to the summed waveforms in each bin of drift time. We extract the measured $D_L$ value from equation (\ref{eqn:diffusionEqn}) using the simulated drift velocity $v_d$ = 1.098 mm/$\mu$s. This simplified sample results in a measured $D_L$ value of 6.30 cm$^2$/s on the collection plane. Compared to the nominal (default) simulation value of 6.40 cm$^2$/s, the measured value is well within the estimated systematic uncertainties, discussed in section \ref{sec:systematics}. The values of $\sigma_t^2(0)$ are extracted from the y-intercept of the linear fit, and their values are close to the expected value of $\sigma_t^2(0) = 1.96$~$\mu \textrm{s}^{2}$. Fit errors on $D_L$ and $\sigma^2_{t}(0)$ are negligible ($<$1\%).

\subsection{Measurement using CRT Data} \label{subsec:crt_measurement}

\begin{figure}
    \centering
    \includegraphics[width=\textwidth]{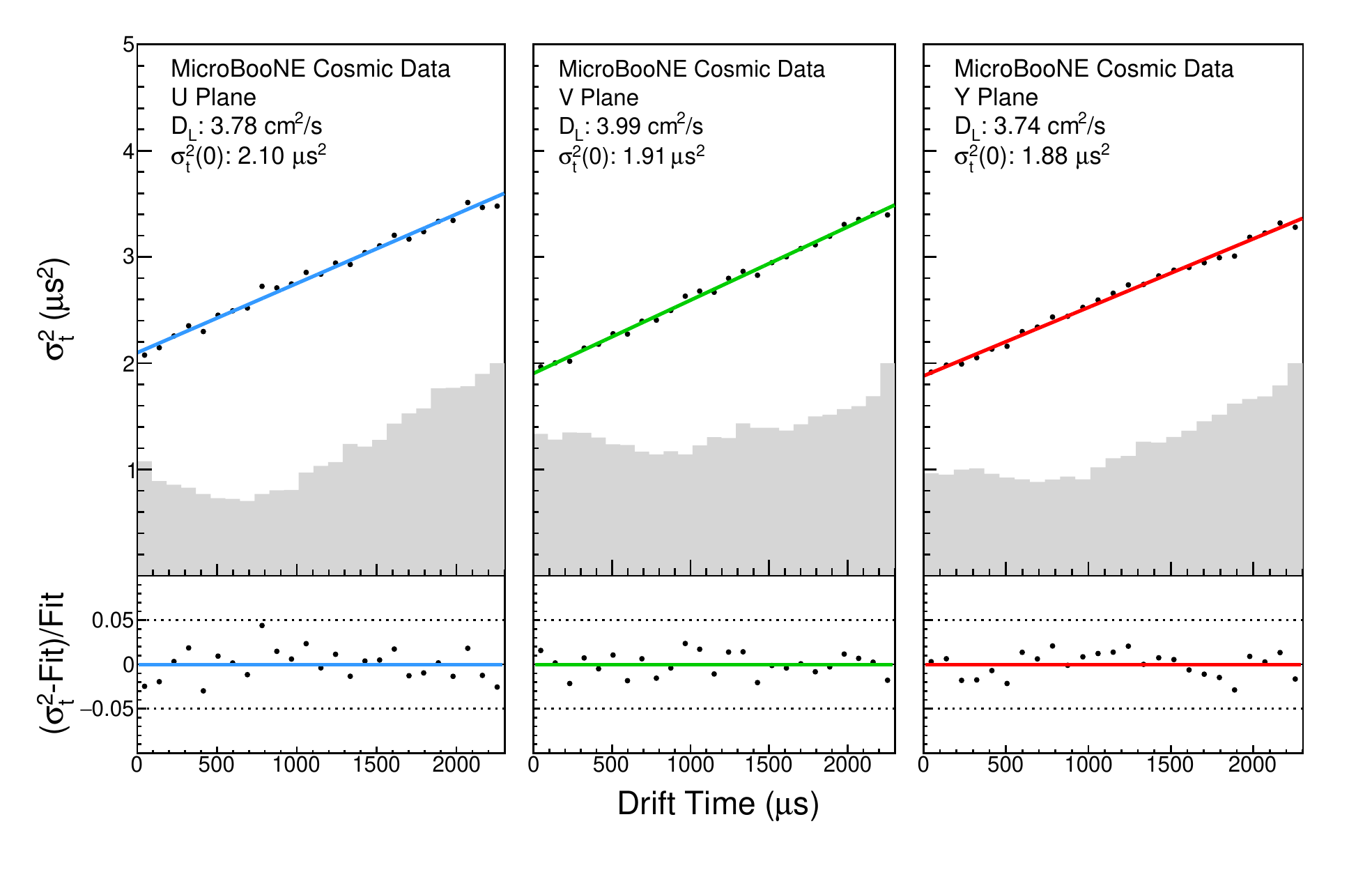}
    \caption{$\sigma_t^2$ versus drift time using MicroBooNE CRT-tagged data. The shaded histograms show the area-normalized distributions of the number of waveforms in each bin. The bottom plots show the fit residuals of each point. The induction planes are used only to estimate systematic uncertainties (see section \ref{sec:systematics}).}
    \label{fig:3planediffresult}
\end{figure}

\begin{figure}
    \centering
    \includegraphics[width=\linewidth]{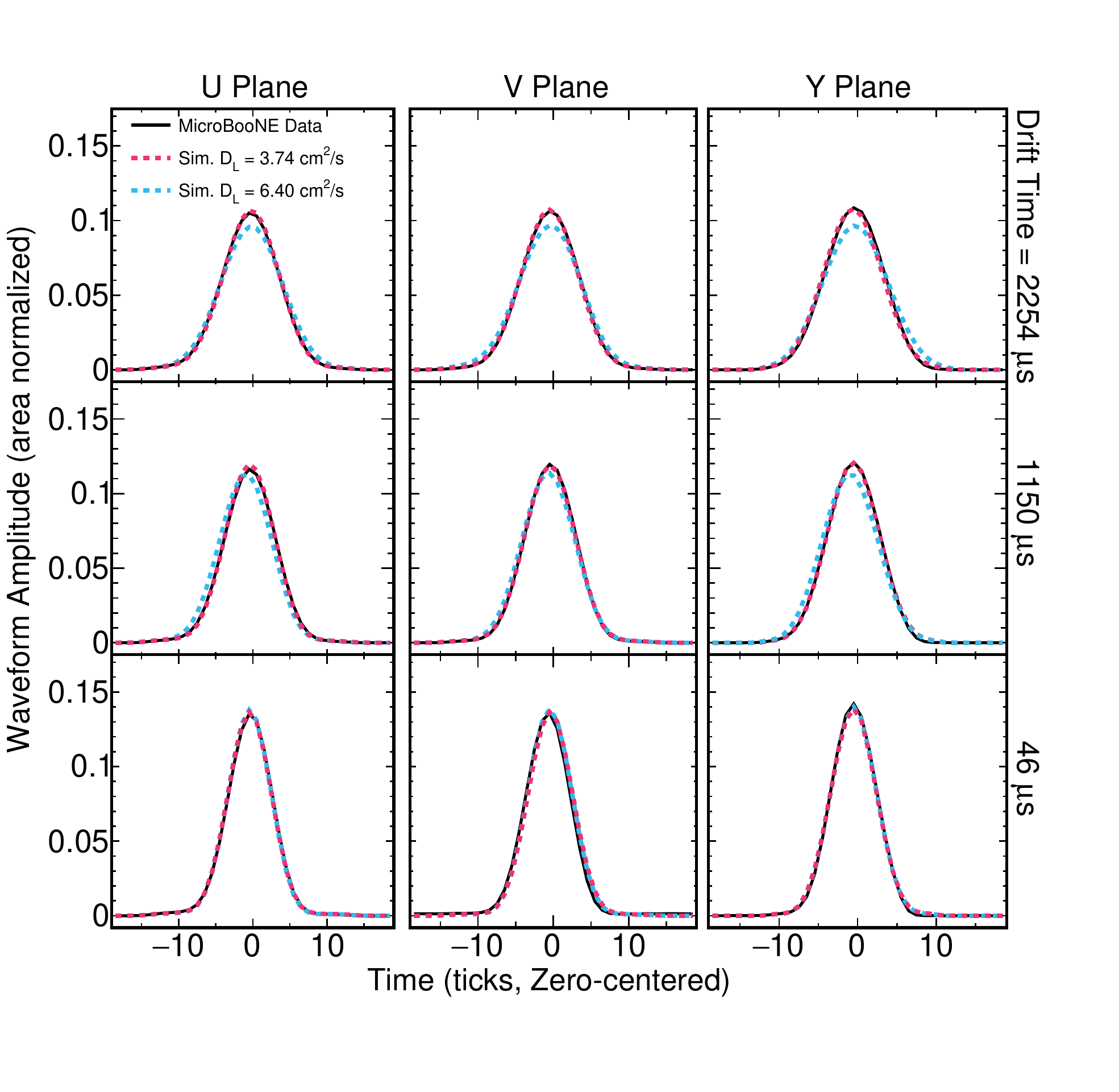}
    \caption{Area-normalized comparisons of summed waveforms at three drift times (46 $\mu$s, $1150 \ \mu$s, $2254 \ \mu$s) for data, and two simulated datasets with different $D_L$ values (6.40 cm$^2$/s and 3.74 cm$^2$/s). To aid in this comparison we have drawn a smooth line through each bin's contents, rather than showing the original digitized data.}
    \label{fig:threeWvfmComparison}
\end{figure}

Figure \ref{fig:3planediffresult} shows the $\sigma_t^2$ versus drift time distribution from which we extract $D_L$ for MicroBooNE data. When using CRT data, the distribution of waveforms peaks near the cathode because of the CRT plane geometry; the CRT plane on the cathode side is nearly twice as large as the anode-side plane. The $D_L$ central value extracted from the slope is 3.74 cm$^2$/s when using collection-plane waveforms. The statistical uncertainties and uncertainties from the fit are negligible. The y-intercept of 1.88 $\mu s^2$ is slightly below the expected 1.96 $\mu$s$^2$. While figure \ref{fig:3planediffresult} shows the fit results on all three wire planes, we choose to quote the value extracted on the collection plane as our measurement. There are two primary reasons for this: 1) the induction planes are known to be more impacted by electronics noise than the collection plane, and 2) the bipolar nature of the induction plane response functions may introduce additional bias in the extracted pulse widths during deconvolution \cite{signal_processing_paper1}. The other wire planes are used for systematic uncertainty studies as described in section \ref{subsec:syst_response}. As a cross-check of this measurement, figure \ref{fig:threeWvfmComparison} shows area-normalized comparisons of summed waveforms between MicroBooNE data and simulated datasets with $D_L = $ 6.40 cm$^2$/s (MicroBooNE nominal) and $D_L = $ 3.74 cm$^2$/s (measured data value). It is clear from these comparisons that the $D_L = 3.74$ cm$^2$/s dataset more closely matches the data waveforms, lending weight to our measurement. Table \ref{tab:measured_dl_values} displays a summary of the results presented in figures \ref{fig:d3planediffresultSim} and \ref{fig:3planediffresult}.  

The extracted $D_L$ value of 3.74 cm$^2$/s differs significantly from the default simulation value of 6.40 cm$^2$/s. Recall figure \ref{fig:dl_summary} which shows a summary of current world data on diffusion. The blue dot-dashed curve shows the parametrization of Li et al. \cite{li2016measurement} while the orange-dashed curve shows the theory prediction of Atrazhev and Timoshkin \cite{Atrazhev:1998}. The default simulation value was extracted from the Li et al. parametrization which is known to be systematically higher than the theory curve. 

\begin{table}[ht]
    \centering
    \begin{tabular}{l|c|c|c}
          & \multicolumn{3}{c}{\textbf{Measured $D_L$ Value (cm$^2$/s)}}\\
          \hline
         \textbf{Sample} & \textbf{U Plane} & \textbf{V Plane} & \textbf{Y Plane}\\
         \hline 
         Simulation ($D_L$ = 6.4 cm$^2$/s) & 6.46 & 6.29 & 6.30 \\
         Data & 3.78 & 3.99 & 3.74 \\
    \end{tabular}
    \caption{Summary of the measured values of $D_L$ from the MicroBooNE data and simulation. The value extracted on the Y plane constitutes our final measurement. The induction planes are used only to estimate systematic uncertainties (see section \ref{subsec:syst_response}). }
    \label{tab:measured_dl_values}
\end{table}

%% file: sections/04_systematics.tex
\section{Evaluation of Systematic Uncertainties}
\label{sec:systematics}

This section describes studies performed to evaluate the total systematic uncertainty on the $D_L$ measurement. While a multitude of effects could potentially bias the measurement, the largest expected systematic effects are due to transverse diffusion, drift velocity variations, and the detector response function modeling. We also considered other possible sources of systematic uncertainty but found them to be sub-dominant. 

\subsection{Transverse Diffusion}\label{subsec:syst_dt}

\begin{figure}[ht]
    \centering
    \includegraphics[width=1.0\textwidth]{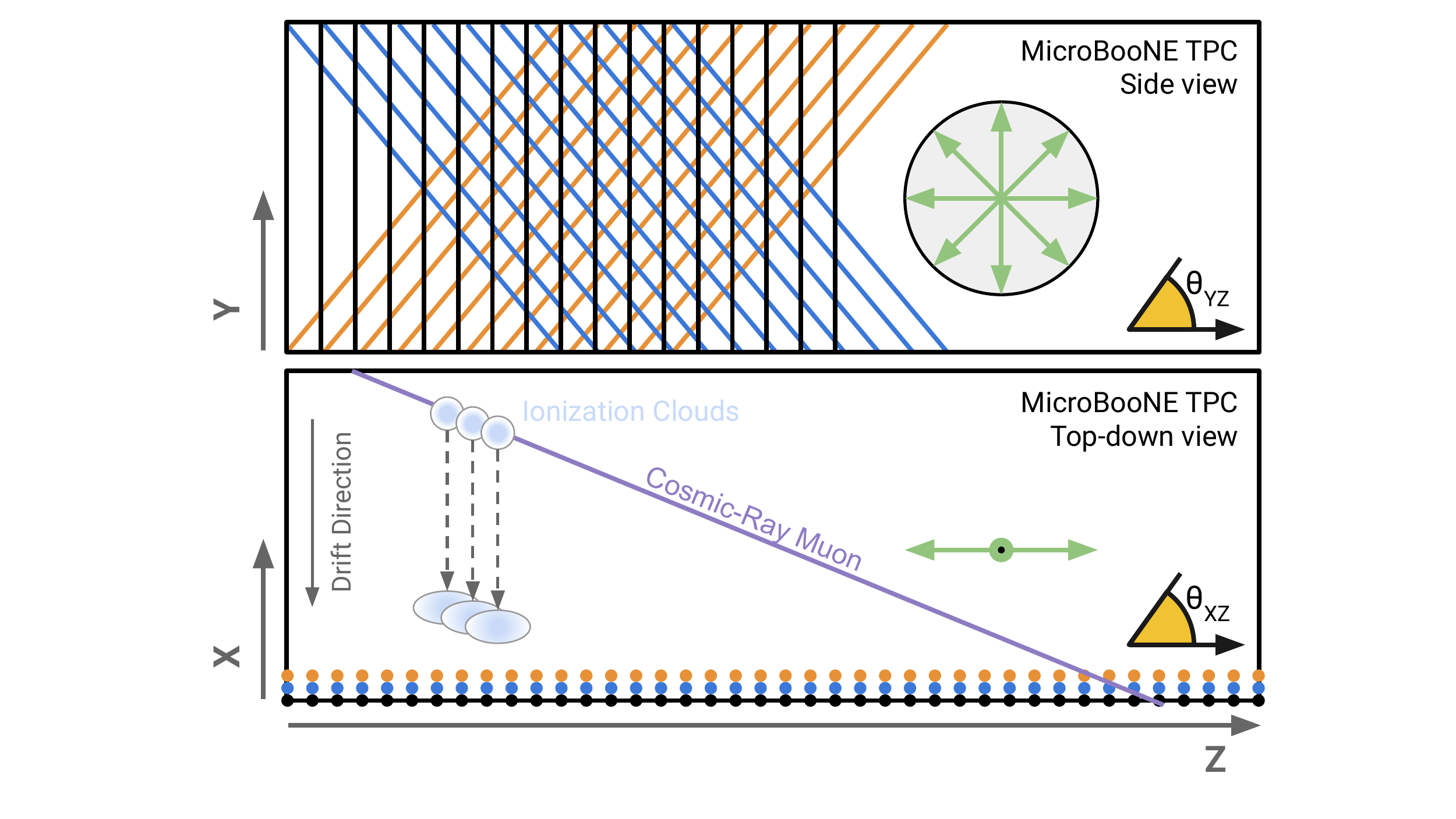}
    \caption{Diagram showing the MicroBooNE coordinate system and wire planes. The beam travels along the $z$-direction, while ionization electrons drift in the decreasing $x$-direction. $y$ denotes the vertical direction. The angles $\theta_{xz}$ and $\theta_{yz}$ denote the angle of a reconstructed object (track or shower) with respect to the beam direction in the $xz$- and $yz$-planes respectively. The top half shows a side view of the TPC through the anode plane, while the bottom half shows a top-down view. The colored lines (dots) on the top (bottom) half represent the three readout wire planes.
    This diagram also depicts the impact of $D_T$ on the $D_L$ measurement. $D_T$ causes electron clouds (light blue gradient) to spread in the $yz$-plane (green arrows) as a function of drift distance. }
    \label{fig:ub_diagram}
\end{figure}

A measured pulse width may have contributions to the pulse width from the effects of $D_T$. Adjacent electron clouds begin to overlap as they spread in the $yz$-plane under the influence of $D_T$, causing additional $\sigma_t$ smearing. Figure \ref{fig:ub_diagram} shows an illustration of this effect. The impact of $D_T$ increases as a function of track $\theta_{xz}$. This motivates the strict angular requirement outlined in section \ref{subsec:track_selection}.

To evaluate a systematic uncertainty on $D_T$, we generate three simulated particle gun samples using the same configuration as the sample described in section \ref{subsec:val_particle_gun} except that we vary the simulated $D_T$ value in each sample. The ratio $D_L/D_T$ can be expressed as 

\begin{equation}
    \frac{D_L}{D_T} = 1 + \frac{E}{\mu(E)}\frac{\partial \mu(E)}{\partial E},
\end{equation}
where $\mu$(E) is the electron mobility as a function of electric field strength \cite{li2016measurement}. Following the parametrization for $\mu(E)$ given in reference \cite{li2016measurement}, we find that that given our measured value of $D_L$ we predict $D_T = 5.85^{+0.62}_{-0.33}$ cm$^2$/s where the uncertainty comes from the uncertainty on the MicroBooNE E-field. We choose $D_T$ variation values of 4.8 cm$^2$/s (down), 5.7 cm$^2$/s (central value), and 7.2 cm$^2$/s (up). These values are scaled linearly by $D_L^{\mathrm{Measured}}/D_L^{\mathrm{Simulated}}$ from the nominal simulated MicroBooNE $D_T$ value and uncertainties, which were designed using the Atrazhev-Timoshkin theory \cite{Atrazhev:1998} and the available world data \cite{Cennini:1995ha, li2016measurement}. 

Table \ref{tab:dt_variation_values} shows the results of running these $D_T$-varied samples through the $D_L$ analysis. The measured $D_L$ central values and $\sigma_0^2$ values show virtually no change when varying $D_T$. We attribute this to the two-dimensional nature of the MicroBooNE deconvolution---which deconvolves the signal in both time and wire space, mitigating the impact of charge spread to neighboring wires---and our stringent requirement on the value of $\theta_{xz}$. We conclude that the uncertainty on $D_T$ does not contribute to the systematic uncertainty on the $D_L$ measurement. 

\begin{table}
    \centering
    \begin{tabular}{c|c|c}
         \textbf{Simulated \bm{$D_T$}} (\textbf{cm}\bm{$^2$}\textbf{/s}) &  \textbf{Measured} \bm{$D_L$} (\textbf{cm}\bm{$^2$}\textbf{/s}) & \textbf{Measured} \bm{$\sigma_0^2$} (\bm{$\mu$}\textbf{s}\bm{$^2$})\\
         \hline
         4.80 & 6.26 & 1.96 \\
         5.70 & 6.26 & 1.97 \\
         7.20 & 6.25 & 1.98 \\
    \end{tabular}
    \caption{Results of $\sigma_t^2$ vs. $t$ fits for simulated muon particle gun samples with $D_T$ varied.}
    \label{tab:dt_variation_values}
\end{table}

\subsection{Drift Velocity}\label{subsec:syst_vd_sce}

Equation (\ref{eqn:diffusionEqn}) shows that $D_L$ is proportional to $v_d^2$, meaning that any uncertainty in $v_d$ could lead to a sizeable systematic uncertainty on $D_L$. MicroBooNE has measured the drift velocity across the active volume of the detector using UV laser and cosmic data \cite{abratenko2020measurement, chen2019_drift_velocity}. Across the anode plane, the drift velocity is not constant due to edge effects near the field cage. To extract $D_L$ from the measured slope in figure \ref{fig:3planediffresult} (using equation (\ref{eqn:diffusionEqn})), we use $v_d=1.076$ mm/$\mu$s, the average value of the measured drift velocity across the anode plane. 

\begin{figure}[ht]
    \centering
    \includegraphics[width=\textwidth]{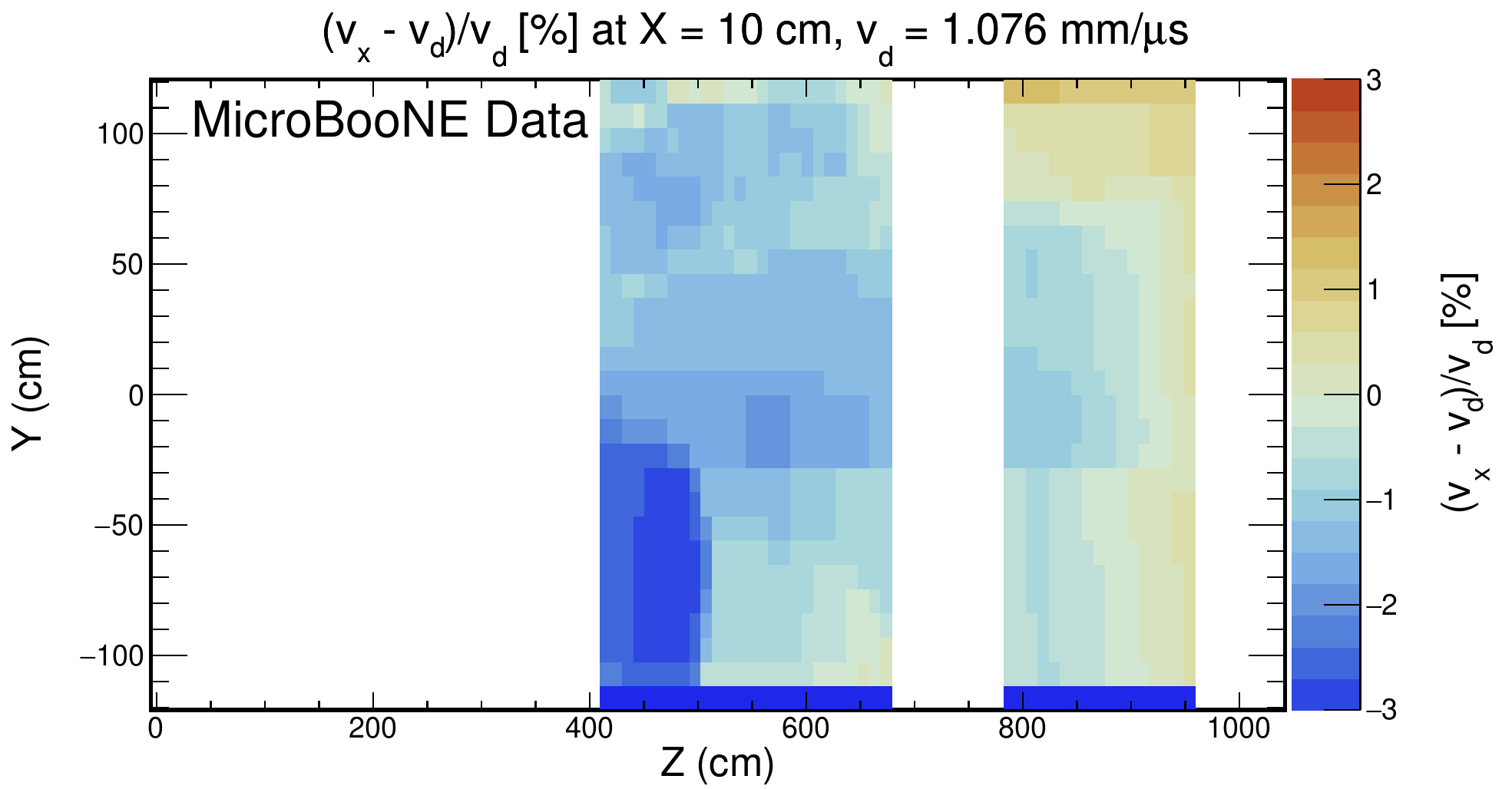}
    \caption{2D distribution of the percent variation of the drift velocity relative to the average drift velocity near the anode, $v_d$ = 1.076 mm/$\mu s$, using the UV laser data map. Here, we have applied the waveform fiducial volume described in section \ref{subsec:waveform_selection}.}
    \label{fig:vmap_laseronly}
\end{figure}

To evaluate a systematic uncertainty on the measurement from the drift velocity, we take 1$\sigma$ variations of $v_d$ near the anode and recalculate $D_L$ using these varied $v_d$ values. Figure \ref{fig:vmap_laseronly} shows a 2D map of the percent variation of $v_d$ with respect to $v_d$ = 1.076 mm/$\mu$s in a $yz$-slice near the anode. The drift velocity values in each bin come from the UV laser data map which was calculated using data from a dedicated calibration run in Summer 2016. Here, we ignore any bins that fall outside the waveform fiducial volume (see section \ref{subsec:waveform_selection}). 

Additional sources of uncertainty on $v_d$ include the statistical and systematic uncertainties on the drift velocity map and cosmic ray flux variations over time. Reference \cite{chen2019_drift_velocity} shows that the uncertainties in the drift velocity map are dominated by statistical errors, but those errors are sub-percent level in each bin in our region of interest. The drift velocity map was calculated using laser data during the Summer of 2016, while the CRT data used in this analysis was taken between October 2017 and March 2018. Time variations of the SCE were studied in reference \cite{abratenko2020measurement} and found to be small compared to the absolute scale of the effect. We therefore conclude that variations in SCE due to cosmic ray flux variations are already accounted for in the drift velocity map.

Because the additional sources of uncertainty discussed in the previous paragraph may be neglected, we set the uncertainty on the drift velocity by looking at the variation of the value of $v_d$ across our fiducial volume. The maximum $v_d$ variation is approximately 3\% in the region near $z = 400$ cm where $y < 0$. However, figure \ref{fig:spacepoint_yz} shows that our selected waveforms fall mostly in the region where $z > 800$ cm. In this region, the drift velocity map shows that the $v_d$ variations are sub-percent level. Because our data mostly lay in the low-uncertainty region, We choose to conservatively apply a $\pm$2\% variation to $v_d$. Varying the anode $v_d$ up and down by 2\% yields variation values of 1.098 mm/$\mu s$ and 1.055 mm/$\mu s$, respectively. This difference covers any impact caused by cosmic ray flux variation and statistical uncertainties in the drift velocity map. The difference also comfortably covers any potential uncertainty from temperature variations over the data taking period, which would result in sub-percent changes in the value of $v_d$. Re-calculating the $D_L$ value shown in figure \ref{fig:3planediffresult} using these variation values, we obtain an asymmetric drift velocity systematic uncertainty of $+3.9$\%, $-4.1$\%. 

\subsection{Detector Response Function}\label{subsec:syst_response}

Equation (\ref{eqn:deconvolution_response}) shows that the MicroBooNE 2D deconvolution depends on the detector response function, $R(\omega)$, as part of the deconvolution kernel. The response function has been validated on each of the three wire planes using MicroBooNE data \cite{signal_processing_paper1}, but small uncertainties on the width of the field response function\footnote{Recall that the response function is itself a convolution of a field response and electronics response.} can have a significant impact on the width of deconvolved waveforms which in turn impacts $D_L$. While the final $D_L$ measurement uses only collection-plane waveforms, we can perform the measurement on each of the three wire planes as shown in figures \ref{fig:d3planediffresultSim} and \ref{fig:3planediffresult}. Since the response function on each plane was tuned independently of the others, we expect some difference in the extracted $D_L$ on each plane. The difference in the measured $D_L$ serves as a conservative estimate of the uncertainty of the wire response tuning method. 

Table \ref{tab:measured_dl_values} shows that the maximum cross-plane difference in $D_L$ is 6.5\% in the data corresponding to the difference between the V and Y planes. We therefore take 6.5\% as the systematic uncertainty on the response function modeling. 

We note here that, by design, applying the deconvolution to the simulation does not perfectly remove the response used during detector simulation \cite{signal_processing_paper1}. Treating this plane-to-plane difference in the simulation analogously to the data, we are able to estimate a simulation-based response-function uncertainty of 2.5\%, which covers the 1.5\% difference observed on the collection plane.

\subsection{Waveform Summation Method}\label{sec:syst_waveform_summation}

The waveform summation technique described in section \ref{subsec:waveform_selection} may introduce additional broadening in the summed waveform. When aligning two waveforms, we can only shift them by integer tick values meaning that the peaks may be misaligned by as much as half a tick. We mitigate this smearing by taking the configuration which minimizes the resultant waveform RMS, but there may still be some residual broadening. 

\begin{figure}[ht]
    \centering
    \begin{subfigure}{.49\textwidth}
        \includegraphics[width=\textwidth]{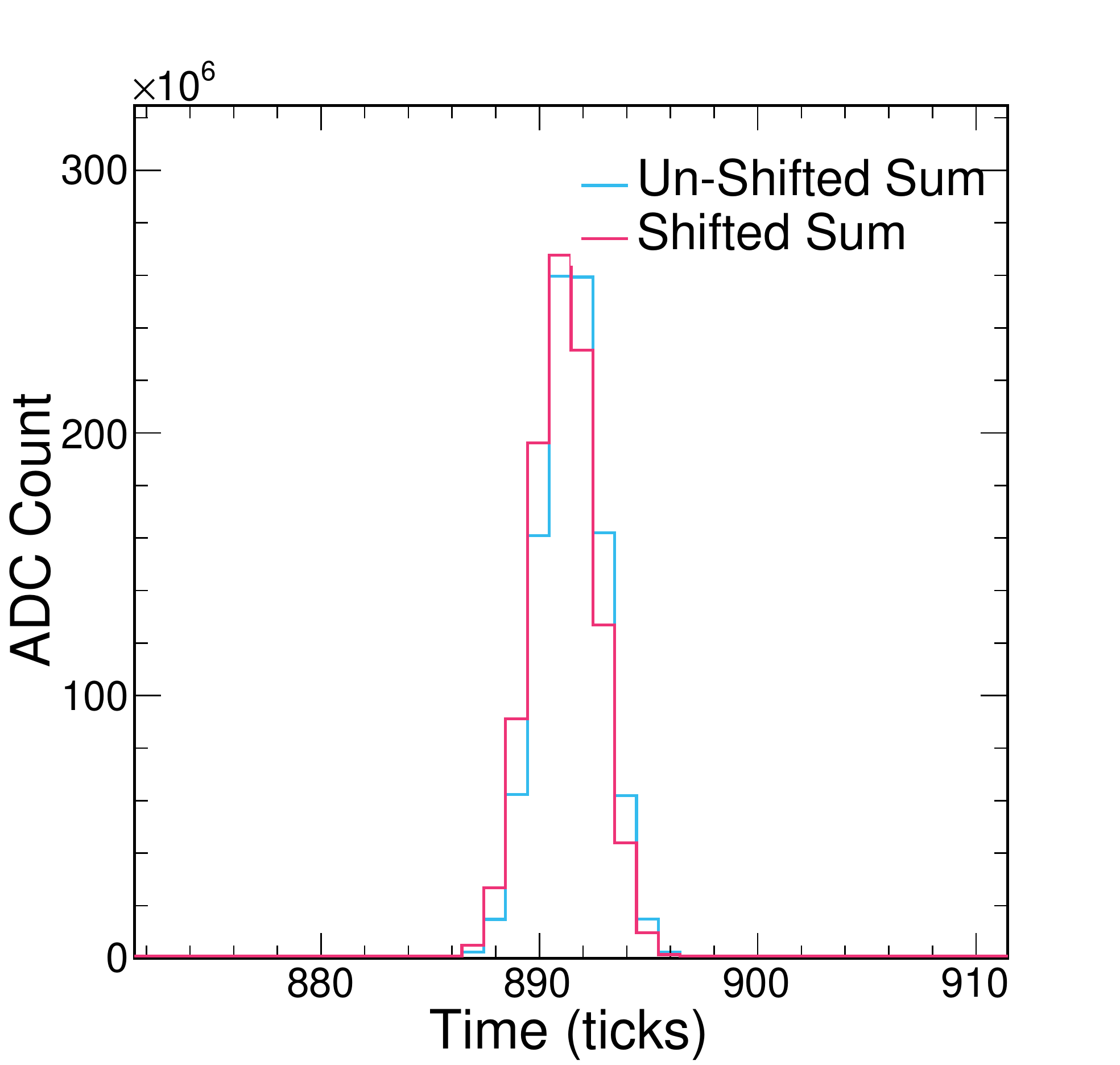}
        \caption{Anode-like}
        \label{fig:waveform_sum_anode}
    \end{subfigure} %
    \begin{subfigure}{.49\textwidth}
        \includegraphics[width=\textwidth]{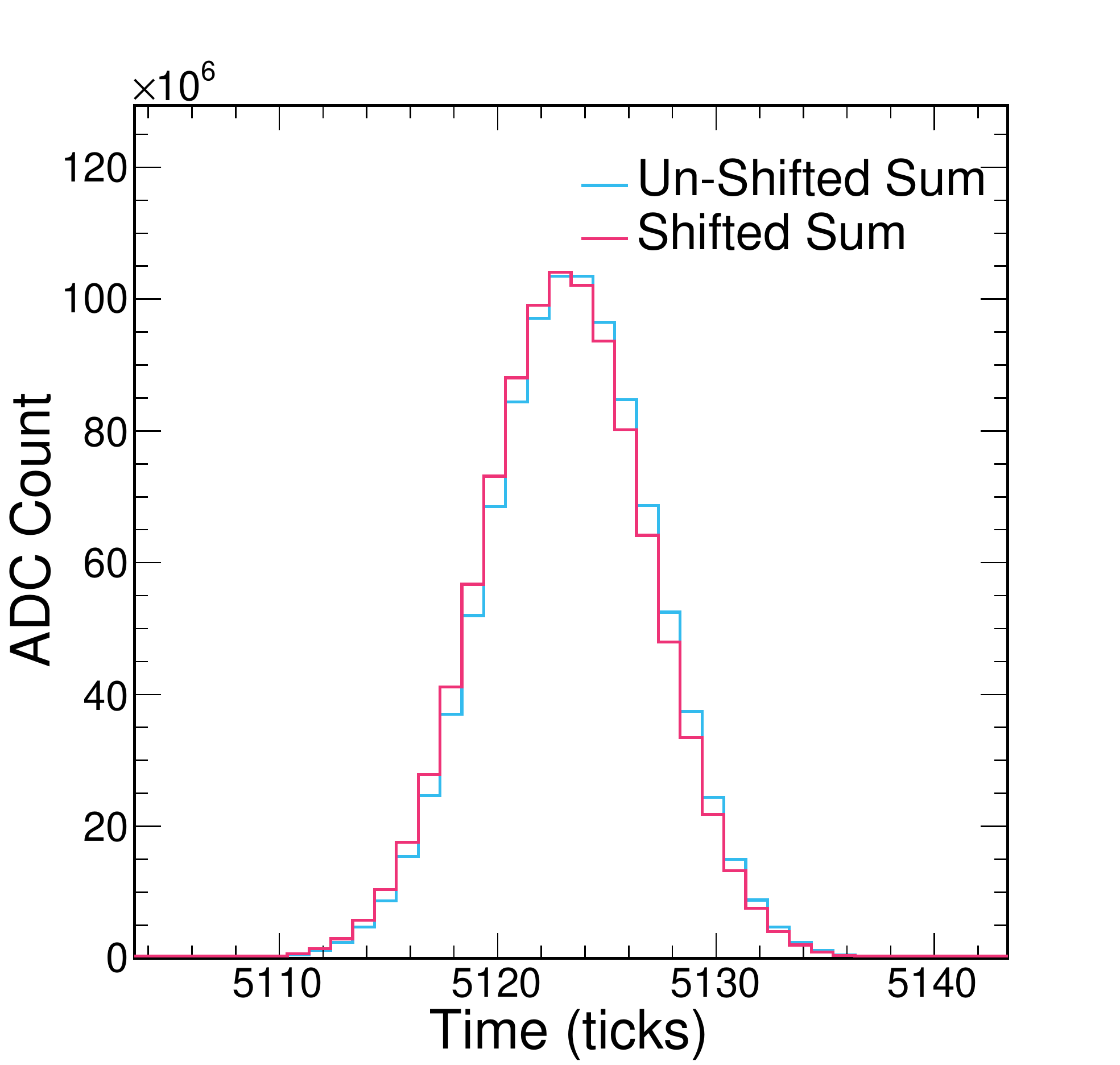}
        \caption{Cathode-like}
        \label{fig:waveform_sum_cathode}
    \end{subfigure} %
    \caption{Results of the study of the waveform summation technique for anode-like (left) and cathode-like (right) Gaussians. }
    \label{fig:waveform_sums_anode_cathode}
\end{figure}

To check the impact of this effect and whether the impact is drift-dependent, we perform a study in which we sum 1000 idealized Gaussian waveforms under different conditions. We start by generating an initial Gaussian whose mean and standard deviation resemble those of waveforms from particle interactions near the anode; here, we chose ``anode-like'' values of $\mu = 891.5$ ticks\footnote{MicroBooNE TPC waveforms are recorded beginning 800 ticks before the trigger time, so the position of the anode is at 800 ticks.} and $\sigma = 1.42$ ticks.\footnote{$\sigma$ here should not be confused with $\sigma_t$, the time width of measured signal pulses.} To simulate the impact of misalignment, we also apply a random shift drawn from a uniform distribution between $-0.5$ and $+0.5$ ticks to the mean of this initial Gaussian. In the control case, we simply add this waveform to itself 1000 times using our waveform summation technique. Then, to simulate the effect of adding misaligned waveforms, we instead add 1000 waveforms with the same $\sigma$ as the initial generated Gaussian, but whose means have been shifted randomly between -0.5 and +0.5 ticks. Any difference in the extracted $\mu$ and $\sigma$ is attributed to the summation technique. We then repeat this study using ``cathode-like'' waveform values of $\mu = 5123.4$ ticks and $\sigma = 3.80$ ticks. Figure \ref{fig:waveform_sums_anode_cathode} and table \ref{tab:waveform_sums} summarize the results of this study. While $\sigma$ does increase slightly in each case, the broadening is consistent at both the anode and the cathode. This may impact our extracted $\sigma_t^2(0)$ but not $D_L$. We repeated this study multiple times to account for different random shifts in the mean of the initial Gaussian and found no significant change in the results, including for cases where the initial Gaussian was shifted by the maximum allowed value ($\pm 0.5$ ticks). We conclude that the waveform summation technique does not introduce a sizeable systematic uncertainty to the $D_L$ measurement. 

\begin{table}
    \centering
    \begin{tabular}{l|c|c}
        & \bm{$\mu$} \textbf{(ticks)} & \bm{$\sigma$} \textbf{(ticks)} \\
        \hline 
         \textbf{Anode-like} & & \\
         \hspace{5mm} Un-shifted & 891.5 $\pm$ 5 $\times 10^{-5}$ 
                      & 1.44 $\pm$ 4 $\times 10^{-5}$ \\
         \hspace{5mm} Shifted & 891.1 $\pm$ 5 $\times 10^{-5}$ 
                      & 1.48 $\pm$ 4 $\times 10^{-5}$ \\
         \textbf{Cathode-like} & & \\
         \hspace{5mm} Un-shifted & 5123.4 $\pm$ 1 $\times 10^{-4}$ & 3.80 $\pm$ 1 $\times 10^{-4}$ \\
         \hspace{5mm} Shifted & 5123.1 $\pm$ 1 $\times 10^{-4}$  & 3.83 $\pm$ 1 $\times 10^{-4}$ \\
    \end{tabular}
    \caption{Results of a toy MC study of waveform summation. ``Un-shifted'' denotes the control case, in which we add the same Gaussian to itself 1000 times, while ``Shifted'' denotes the case in which each added waveform has its mean randomly shifted before addition.}
    \label{tab:waveform_sums}
\end{table}

\subsection{Summary and Other Systematic Uncertainties}

Other systematic uncertainties that may impact the $D_L$ measurement include microphysics effects that are either drift-dependent or field-dependent, particularly SCE, electron-ion recombination, and electron attenuation. For SCE, the size of the electron cloud when it arrives at the anode wire plane depends only on the amount of time that has elapsed since the electrons were ionized. We measure this time directly by using the $t_0$ extracted from CRT information meaning that the measurement is not biased by the presence of space charge. Thus, the measured slope of the line in figure \ref{fig:3planediffresult} has no systematic uncertainty due to space charge. The strength of electron-ion recombination changes with the electric field, but, for MicroBooNE E-field fluctuations, this effect is small \cite{adams2020calibration}. Moreover, the impact of the recombination systematic uncertainty on collected charge is much smaller than the impact of statistical Landau fluctuations in the density of ionization electron clouds. As for electron attenuation, the measured electron lifetime in MicroBooNE \cite{meddage2017electron} is 18 ms. The maximum drift for a single electron is 2.3 ms meaning that charge attenuation in MicroBooNE is minimal, and this is due to the extremely high argon purity in the TPC. We conclude that both electron recombination and attenuation do not contribute to the systematic uncertainty on $D_L$. 

Table \ref{tab:systematics} summarizes the $D_L$ systematic uncertainties. The two dominant systematic uncertainties come from the uncertainties on the response function modeling and the drift velocity. We have considered many other potential sources of systematic uncertainties but found them to be sub-dominant. We assume that the individual systematic uncertainties are uncorrelated and add them in quadrature to obtain the total systematic uncertainty of $+$7.6\%, $-$7.7.\%. This results in our final measurement from the MicroBooNE data of $D_L = 3.74^{+0.28}_{-0.29}$ cm$^2$/s. 

\begin{table}[!ht]
\makebox[\textwidth]{
	\begin{tabular}{ l | r }
   		\textbf{Systematic} & \textbf{Value} \\
        \hline
        Response Function       & $6.5$\%  \\
  		Drift Velocity          & +3.9\%, -4.1\% \\
  		$D_T$                   & $<1$\% \\
  		Waveform Summation      & $<1$\% \\
        Noise and microphysics  & $<1$\% \\
        %Run-to-run variations   & $<1$\% \\
        \hline
        \textbf{Total}          & +7.6\%, -7.7\%  
	\end{tabular}}
\caption{Summary of systematic uncertainties on the $D_L$ measurement. The total uncertainty assumes that the systematic uncertainties are uncorrelated.}
\label{tab:systematics}
\end{table}

%% file: sections/07_discussion.tex
\section{Discussion}

\begin{figure}
    \centering
    \includegraphics[width=0.8\linewidth]{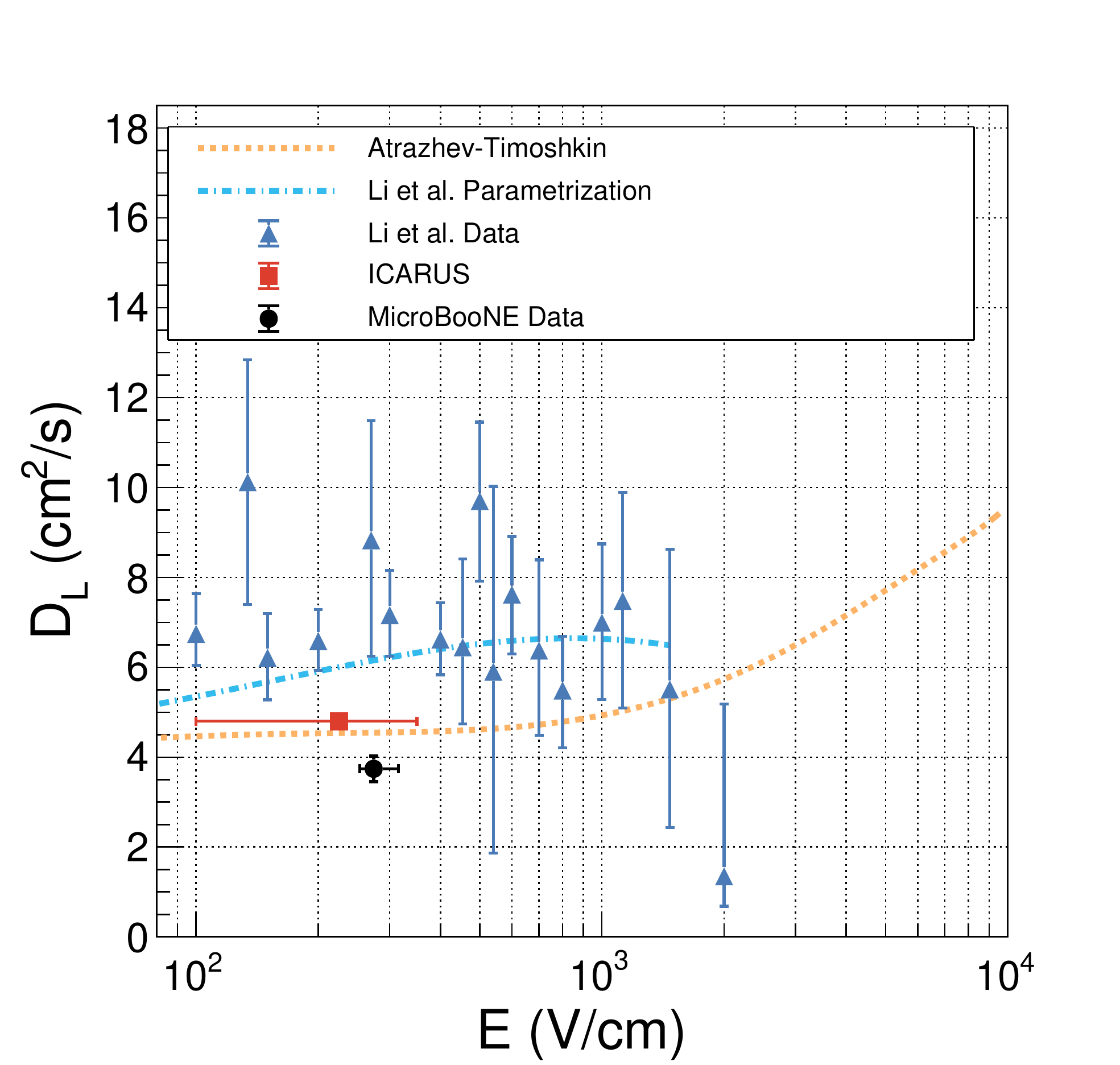}
    \caption{Comparison of the MicroBooNE result with world data for longitudinal electron diffusion in liquid argon. The orange-dashed curve shows the theory prediction from Atrazhev-Timoshkin \cite{Atrazhev:1998}, the blue dot-dashed curve shows the parametrization from Li et al. \cite{li2016measurement}, and the red and dark blue points show the ICARUS \cite{Cennini:1995ha} and Li et al. measurements, respectively. Details of this plot can be found in Appendix \ref{app:diffusion_plot_details}. Note that the ICARUS error bars ($\pm$ 0.2 cm$^2$/s) are covered by the data point.}
    \label{fig:dl_summary}
\end{figure}

Our measured central value of 3.74 cm$^2$/s is more consistent with the Atrazhev-Timoshkin curve than with the Li et al. parametrization. It is of note that the published ICARUS measurement is also in better agreement with the Atrazhev-Timoshkin curve than the Li et al. parametrization, as shown in figure \ref{fig:dl_summary}. We note that the Atrazhev-Timoshkin curve presented in this figure requires an interpolation between the low and high E-field regions. The details of this interpolation can be found in appendix \ref{app:diffusion_plot_details}. The tension between available models and measurements has historically motivated conservative systematic uncertainties on $D_L$ when generating systematically fluctuated simulated samples. This has translated to larger systematic uncertainties for high-level physics analyses at MicroBooNE. At present, the cause of tension among $D_L$ measurements is unknown. Li et al. performed their measurements using a gridded drift cell, similar to historical measurements performed in gaseous media \cite{hunter1986electron, Kusano_2012}, with a maximum drift distance of 60 mm. They note the possibility of underestimating the impact of Coulomb repulsion among the drifting electrons, which they calculated using an approximate model described in reference \cite{Shibamura:1979phx}. Based on their calculations, Li et al. chose not to apply a correction for this effect. ICARUS, however, concluded that this effect contributes significantly to their measured value when using the same model \cite{Cennini:1995ha}. Following the prescription in reference \cite{Shibamura:1979phx}, we conclude that Coulomb repulsion does not affect our measurement. Any effect in the direction perpendicular to the field would not change the measured value of $D_L$, and the effect in the direction parallel to the field is negligible. Further, any contribution to the pulse width from Coulomb repulsion goes as $\sqrt[3]{t}$, which would result in a non-linearity when plotting $\sigma_t^2$ as a function of $t$. We do not observe this in our data (figure \ref{fig:3planediffresult}), indicating that any contribution to the width must be small. Further measurements in LArTPCs are needed in order to resolve this tension. The potential for using diffusion to $t_0$-tag single waveforms has been investigated and is presented in Appendix \ref{app:t0tagging}.

%% file: sections/05_conclusions.tex
\section{Conclusions}\label{sec:conclusions}

We report a measurement of the effective longitudinal electron diffusion coefficient of $D_L = 3.74^{+0.28}_{-0.29}$ cm$^2$/s at an E-field of 273.9 V/cm. This represents the first measurement in a large-scale (85 tonne) LArTPC. Figure \ref{fig:dl_summary} shows the measured $D_L$ value in MicroBooNE as it compares to the Li et al. parametrization, the Atrazhev-Timoshkin theory curve, and the available data from ICARUS and Li et al. The vertical error bars correspond to systematic uncertainties on $D_L$, while the horizontal error bars account for the maximum E-field variation values of $273.9_{-8\%}^{+12\%}$. The MicroBooNE $D_L$ value sits slightly below the theory curve even when including systematic uncertainties, but it should be noted that this curve is ill-defined for E-fields greater than zero and below $\sim$1 kV/cm. We used an interpolation in that region, the details of which are described in Appendix \ref{app:diffusion_plot_details}. Our measurement is in better agreement with the ICARUS measurement and the Atrazhev-Timoshkin prediction than the measurement and parametrization of Li et al.

%% file: sections/06_acknowledgements.tex
\section{Acknowledgements}
This document was prepared by the MicroBooNE collaboration using the resources of the Fermi National Accelerator Laboratory (Fermilab), a U.S. Department of Energy, Office of Science, HEP User Facility. Fermilab is managed by Fermi Research Alliance, LLC (FRA), acting under Contract No. DE-AC02-07CH11359.  MicroBooNE is supported by the following: the U.S. Department of Energy, Office of Science, Offices of High Energy Physics and Nuclear Physics; the U.S. National Science Foundation; the Swiss National Science Foundation; the Science and Technology Facilities Council (STFC), part of the United Kingdom Research and Innovation; and The Royal Society (United Kingdom).  Additional support for the laser calibration system and cosmic ray tagger was provided by the Albert Einstein Center for Fundamental Physics, Bern, Switzerland.

%% file: appendices/a01_diffusionplotdetails.tex
\section{Diffusion World Data Comparison Plot Details}\label{app:diffusion_plot_details}

This appendix is dedicated to a description of the production of figure \ref{fig:dl_summary}. Both Atrazhev-Timoshkin \cite{Atrazhev:1998} and Li et al. \cite{li2016measurement} present their results in terms of longitudinal electron energy, $\epsilon_L$, while the ICARUS results \cite{Cennini:1995ha} and those results presented in this work are presented in terms of $D_L$. A conversion must be applied in order to directly compare the results. The $\epsilon_L$ parameter is related to $D_L$ via the generalized Einstein-Smoluchowski relation \cite{wannier1953motion, robson1972thermodynamic}, 

\begin{equation}
    D_L = \frac{\mu(E)\epsilon_{L}}{e},
    \label{eq:el_to_dl}
\end{equation}
where $e$ is the electron charge and $\mu(E)$ is the electron mobility.

\subsection{Details on Datasets and Theory Curves}
\label{subsec:dataset_details}
\subsubsection{Atrazhev-Timoshkin Theory}
The Atrazhev-Timoshkin theory is described in reference \cite{Atrazhev:1998}. It is noted that for field strengths $>10^{3}$ V/cm $\epsilon_L$ can be described as

\begin{equation}
    \epsilon_{L} = 0.5 \epsilon_{T},
    \label{eq:el_atrazhev}
\end{equation}
where $\epsilon_{T}$ is the transverse electron energy which is given by

\begin{equation}
    \epsilon_{T} = 0.8 T (E/E_{h}).
\end{equation}
Reference \cite{Atrazhev:1998} also notes that below the boundary field strength, $E_{h}$, $\epsilon_L = T$, the temperature of the liquid argon. There is no description of $\epsilon_{L}$ in the transition region from low-E to E $>10^3$ V/cm, and so we must interpolate between the two. This means that the Atrazhev-Timoshkin prediction for both $\epsilon_L$ and $D_L$ should be taken to have large uncertainties at both the MicroBooNE electric field (273.9 V/cm) and also at the planned electric field of other future LArTPCs (500 V/cm).

To interpolate between the two well defined regions, we fit a fourth degree polynomial between points below 10 V/cm, which are set to $T=7.67\times10^{-3}$ eV (89 K) and the points above 1200 V/cm which follow equation (\ref{eq:el_atrazhev}). The resulting functional form is

\begin{align*}
    \epsilon_L = & 7.67\times10^{-3}~\mathrm{[eV]}  \\
                 & + 1.39\times10^{-5}~\mathrm{[eV / (V/cm)]} E\\
                 & + 2.19\times10^{-9}~\mathrm{[eV / (V/cm)^{2}]} E^{2} \\
                 & - 2.69\times10^{-13}~\mathrm{[eV / (V/cm)^{3}]} E^{3}\\
                 & + 1.15\times10^{-17}~\mathrm{[eV / (V/cm)^{4}]} E^{4}.
\end{align*}

\subsubsection{Treatment of the Parametrization and Data of Li et al.}

Reference \cite{li2016measurement} constructs an $\epsilon_L$ parametrization based on the world data (see figure 11 of \cite{li2016measurement}). At low electric field strengths, this parametrization is dominated by their own data and by the ICARUS data. The functional form and the parameter values are provided in reference \cite{li2016measurement}. The Li et al. data points are estimated from figure 11 of \cite{li2016measurement}. 

\subsubsection{Treatment of ICARUS Data}

The ICARUS data is taken directly from \cite{Cennini:1995ha} and are scaled linearly from 92~K to 89~K. 

\subsection{Comparison of MicroBooNE Result with World Data}

After treating the available data and the Atrazhev-Timoshkin curve as outlined in section \ref{subsec:dataset_details}, we can choose to present the data in terms of $D_L$ or $\epsilon_L$ by converting between the two parameters using equation (\ref{eq:el_to_dl}) where $e$ is taken to be 1. For $\mu$, we use a second parametrization from reference \cite{li2016measurement} which is shown in their figure 10 to have excellent agreement with world data.

\begin{figure}[ht]
    \centering
    \includegraphics[width=0.8\linewidth]{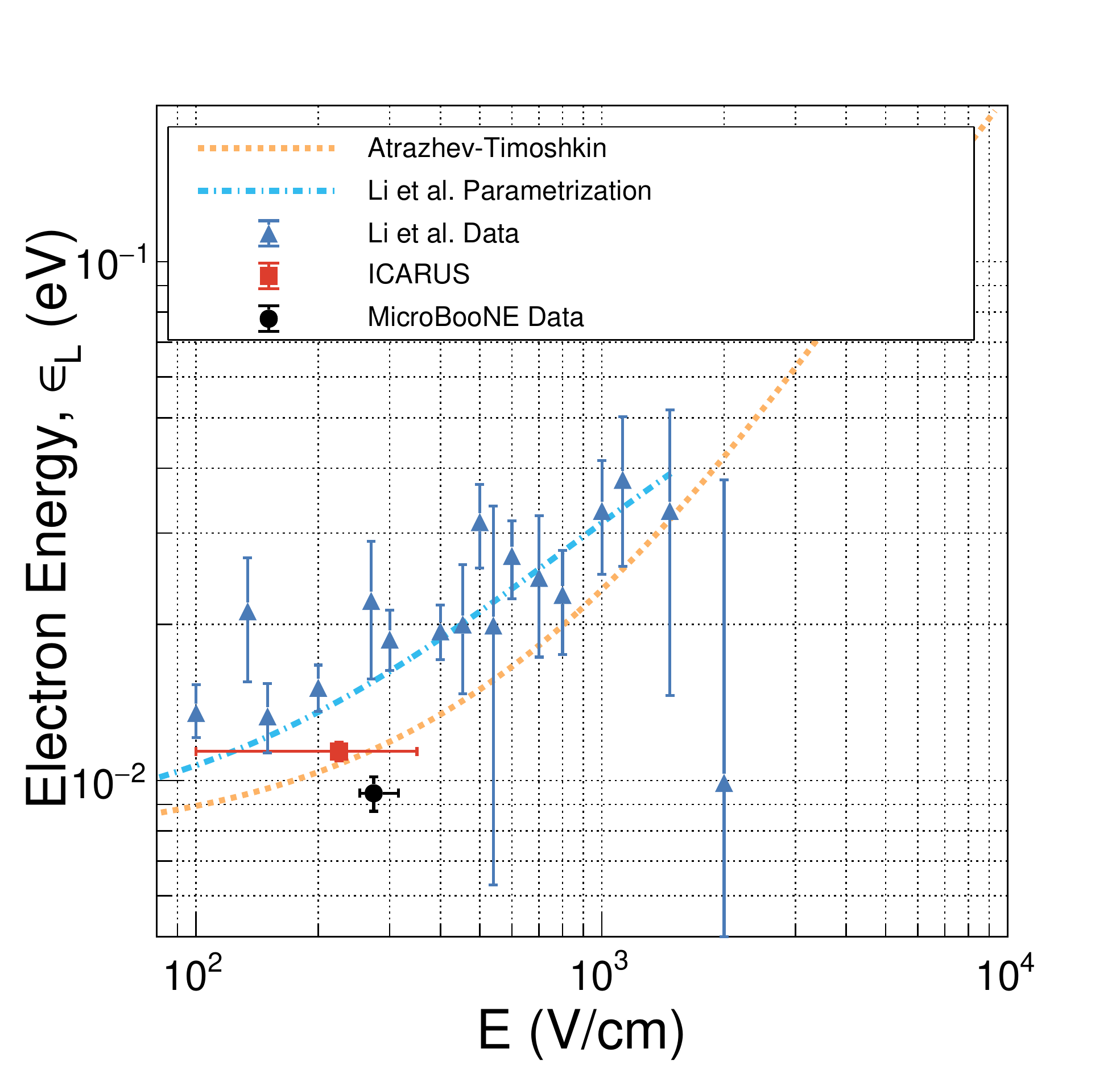}
    \caption{Comparison of the MicroBooNE result with world data for $\epsilon_L$, along with the Atrazhev-Timoshkin theory curve and the Li et al. parametrization.}
    \label{fig:el_summary}
\end{figure}

The primary results of this work are presented in terms of $D_L$ (figure \ref{fig:dl_summary}). For fullness, we also present in figure \ref{fig:el_summary} the data in terms of $\epsilon_L$.

%% file: appendices/ao2_t0taggingpotential.tex
\section{Potential for Tagging \texorpdfstring{$t_0$}{t0} Using Diffusion}
\label{app:t0tagging}

\begin{figure}[t]
    \centering
    \includegraphics[width=\linewidth]{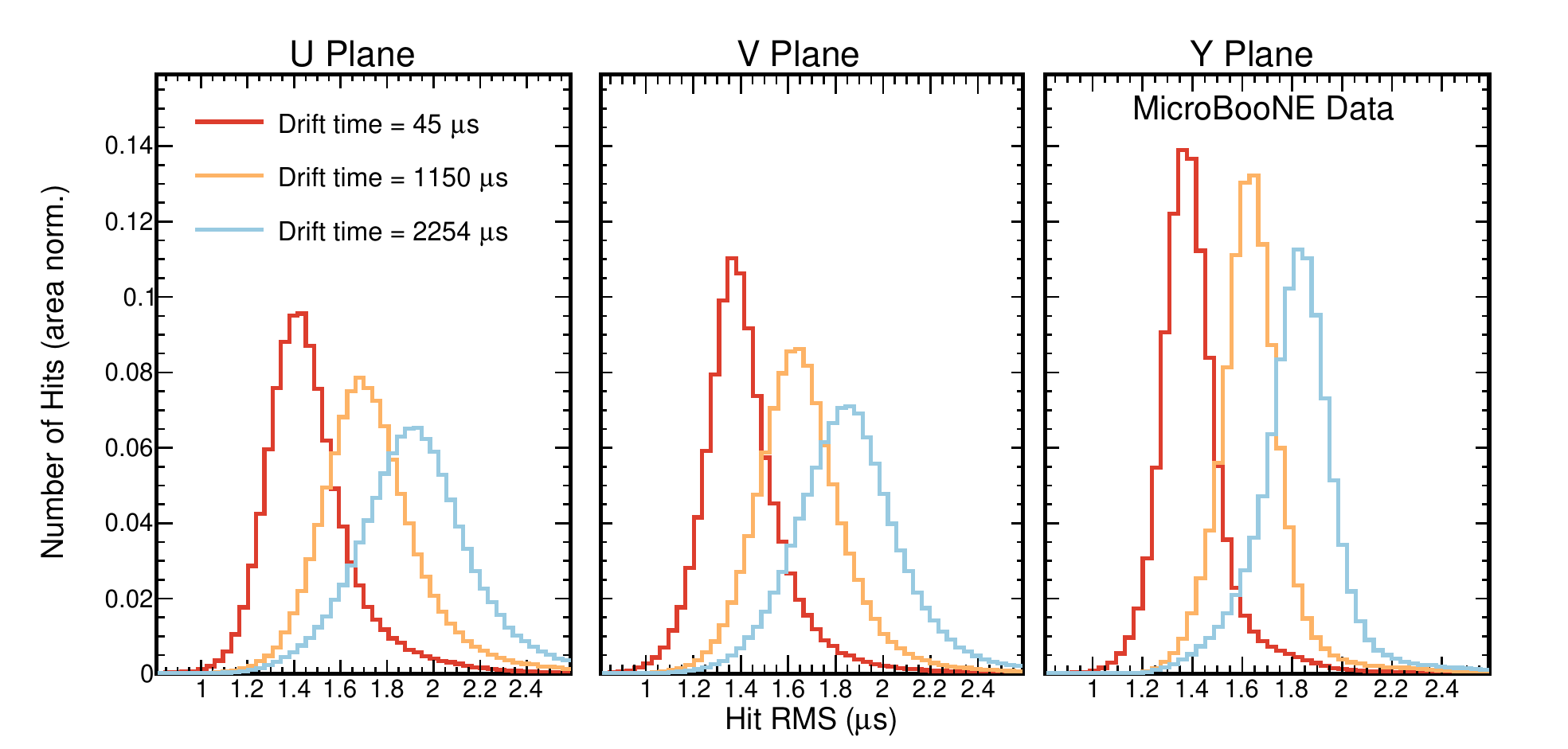}
    \caption{One-dimensional comparisons of the area-normalized hit RMS distribution on the three wire planes for drift times of 45 $\mu$s, 1150 $\mu$s, and 2254 $\mu$s.}
    \label{fig:onedhitrms}
\end{figure}

The potential for $t_0$-tagging using diffusion has been investigated in reference \cite{Warburton:2017} where many hits along a single track are considered in order to reconstruct a $t_0$ for that track. Recently, this method has gained some attention in the context of $t_0$-tagging individual energy depositions. The feasibility of performing $t_0$-tagging for individual energy depositions using this method is dependent on the spread of the hit RMS values for each drift time. This is shown in figure ~\ref{subfig:dynamic_sigma_precut}, where each bin in drift time has a wide range of allowed hit widths. In addition, comparisons of the hit RMS distributions at drift times of 45 $\mu$s, 1150 $\mu$s, and 2254 $\mu$s  are shown in figure \ref{fig:onedhitrms}. Each of these plots uses the nominal angular selection of this analysis, $\theta_{xz} < 6^{\circ}$, meaning this should be comparable to a point source. Figure \ref{fig:onedhitrms} shows that the spread in the hit RMS is relatively wide on all three planes. In order to boost the success rate of tagging the $t_0$ of individual energy depositions, one may imagine performing charge matching across planes in order to obtain three hits rather than one; however, statistical fluctuations in electron transport are likely much larger than any plane-to-plane differences that might be present in a given LArTPC. The collection plane has the narrowest hit RMS distributions and therefore should be the most promising for $t_0$ tagging individual waveforms, and so we focus the rest of this appendix there. 

\begin{figure}[!ht]
    \centering
    \includegraphics[width=\linewidth]{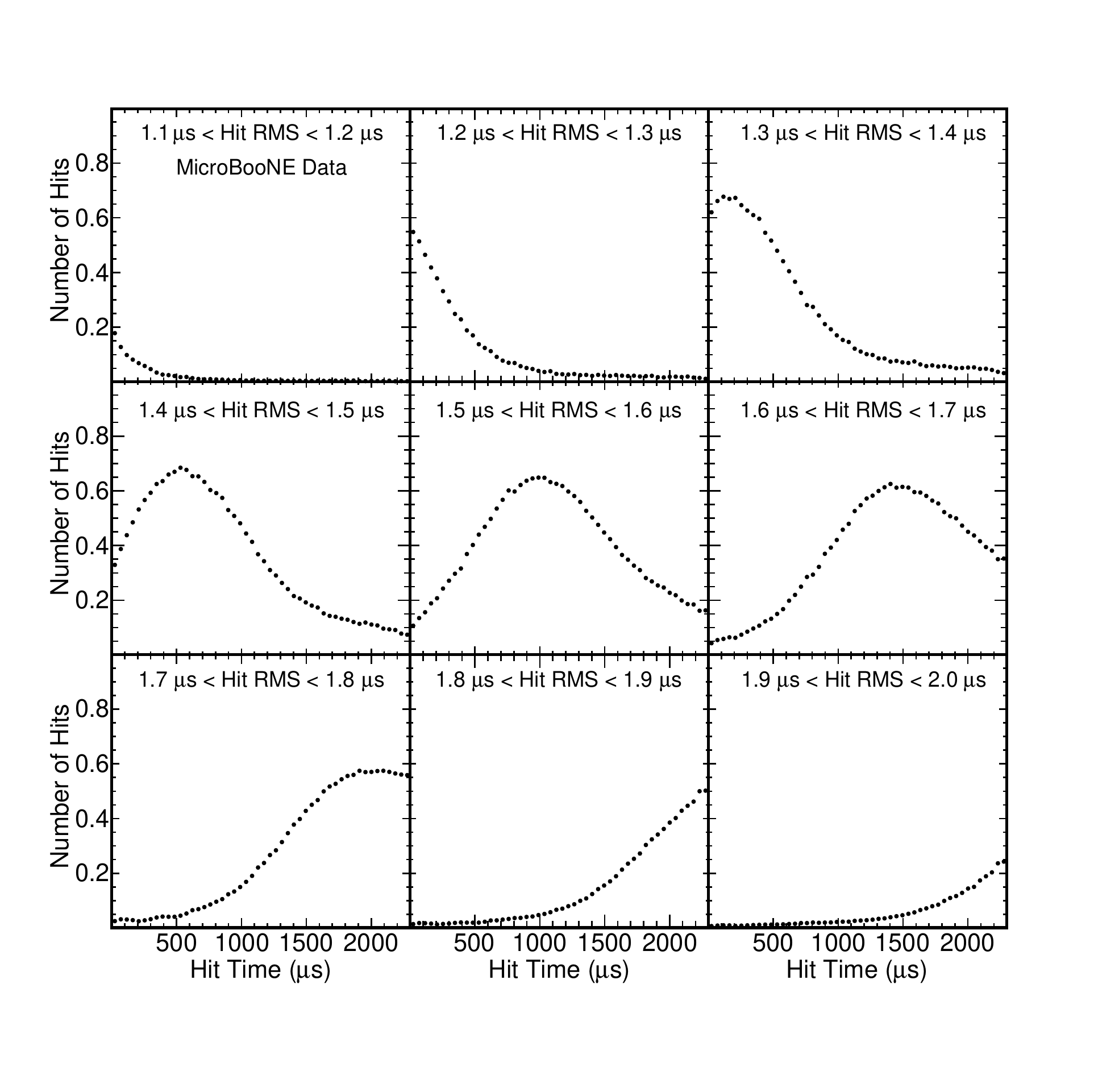}
    \caption{Distribution of hit times for different slices in hit RMS on the collection plane.}
    \label{fig:hit_time_per_hit_rms}
\end{figure}

The ability to $t_0$-tag a single energy deposition accurately relies on each hit RMS value corresponding to a tight distribution of possible drift times. The wider the distribution of possible drift times, the less accurately the $t_0$ can be measured. Figure \ref{fig:hit_time_per_hit_rms} shows the distribution of hit times on the collection plane for hits in 0.1 $\mu$s bins of hit width from zero to the maximum drift time in the MicroBooNE TPC, 2300 $\mu$s. For each bin of hit RMS, the range of drift times spans the entire 0-2300 $\mu$s region. To make a more quantitative statement, we fit a Gaussian functional form around the peak of the $1.5$~$\mu\mathrm{s} < $ Hit RMS $ < 1.6$~$\mu\mathrm{s}$ plot, from which we can estimate a 1 $\sigma$ uncertainty of approximately $\pm$560 $\mu$s. However, we caution that the distribution is relatively non-Gaussian and this should be taken as a lower bound on the resolution. We also note that the resolution is likely larger for hits with larger drift times because the hit width is proportional to $\sqrt{t}$ and the width changes more slowly for longer drifts. The first 10-kTon module of the DUNE Far Detector is planned to have a drift distance of 3.6 m with a drift field of 500 V/cm resulting in a maximum drift time of 2.25 ms. The data presented in this work cover this region of drift time and the field dependence of $D_L$ is negligible (figure \ref{fig:dl_summary}), making this measurement relevant for the DUNE Far Detector.

It is clear that even using the collection plane, which is expected to out-perform the induction planes, there remain significant hurdles to overcome. The measured central value of $D_L$ combined with statistical fluctuations from the diffusion process means that $t_0$ tagging of individual energy depositions using hit RMS alone will result in poor time resolution. Combination of the hit RMS with other variables has not been investigated in this work. Application of this technique to charged particle tracks which are reconstructed from energy depositions on many readout channels remains an intriguing possibility, as the statistical fluctuations will average out as the number of hits increases.

\begin{figure}[t]
    \centering
    \includegraphics[width=0.8\linewidth]{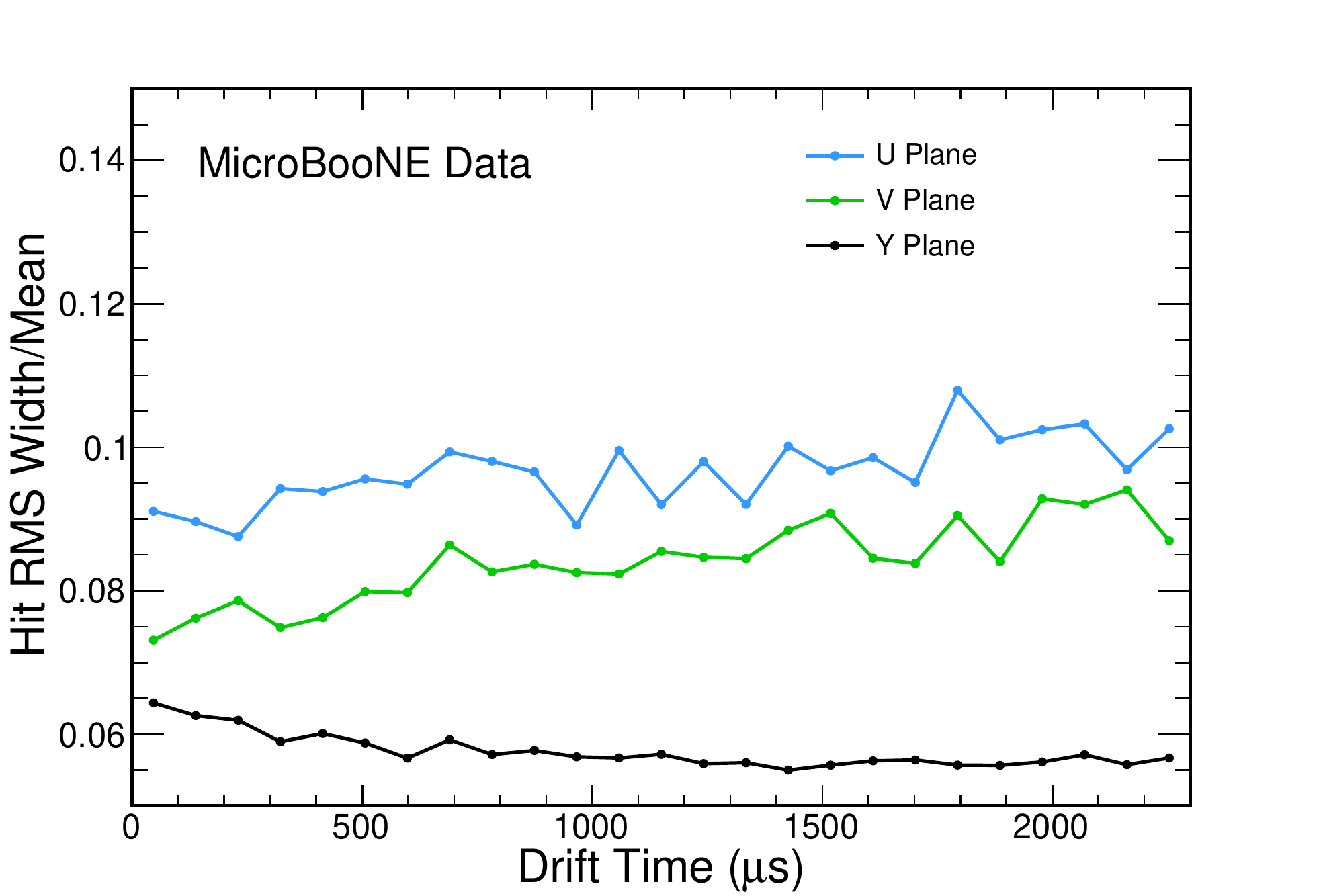}
    \caption{Distribution of the width of the hit RMS distribution over the mean of that distribution for the U, V and Y planes.}
    \label{fig:hit_width_over_mean}
\end{figure}

\begin{figure}[t]
    \centering
    \includegraphics[width=0.6\linewidth]{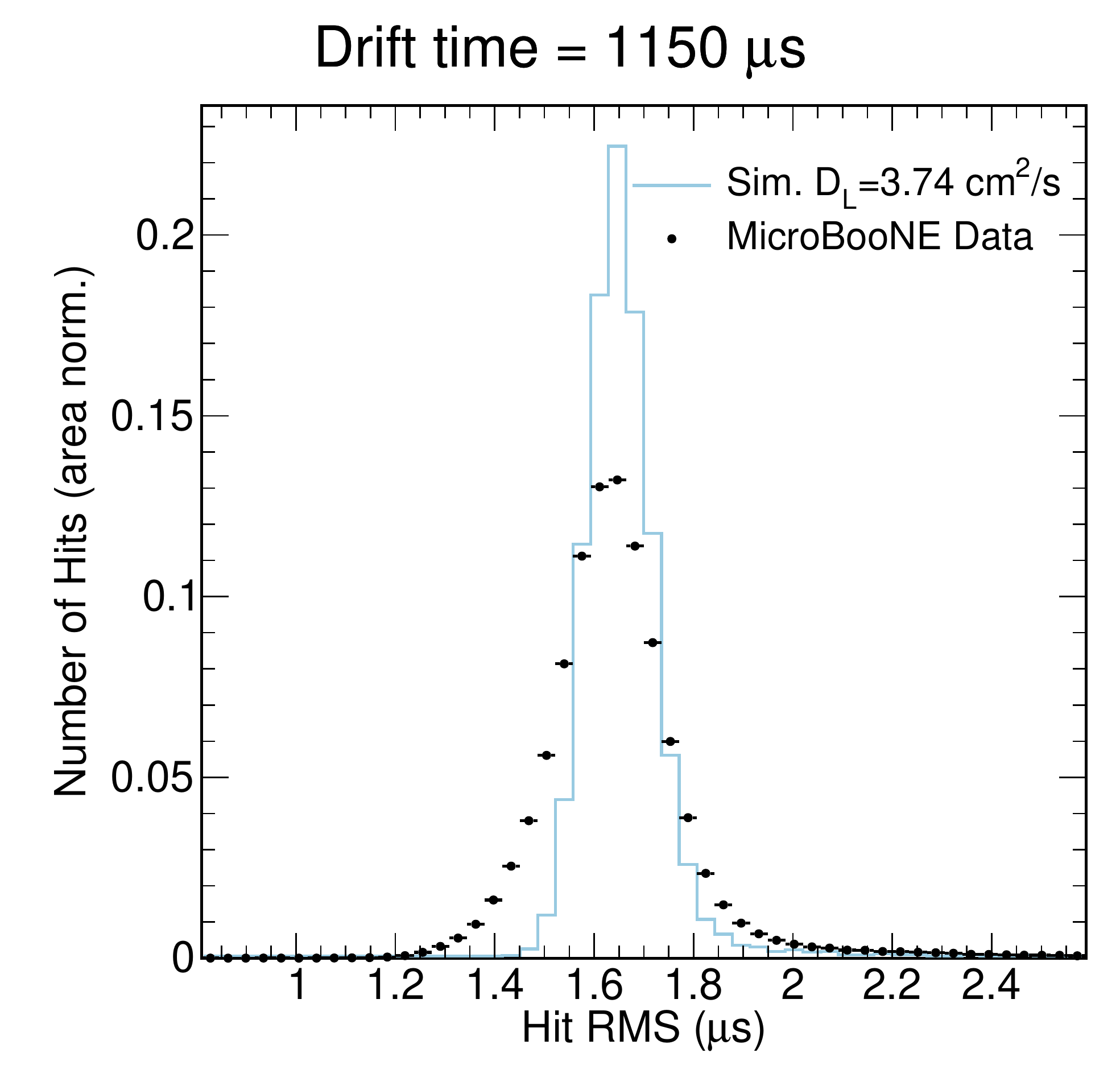}
    \caption{Comparison of data and simulation for the hit RMS distribution in the center of the TPC, around drift time = 1150 $\mu$s.}
    \label{fig:hit_rms_center_bin}
\end{figure}

To aid in making predictions for future long-drift detectors, we make the observation that for drift times above $\sim 1000~\mu$s the ratio of the width to the hit RMS distribution with the mean of the hit RMS distributions is approximately constant at 0.056 (figure \ref{fig:hit_width_over_mean}). This relationship does not appear to hold for the induction planes. We provide this extrapolation for use with other LArTPCs, but we emphasize that this is not a substitution for a full analysis with a dedicated simulation. Such an endeavor demands more precision from simulations than has been required to date. For example, we have noted that the distribution of the hit RMS for a given drift time tends to be narrower in our simulations than in our data (figure \ref{fig:hit_rms_center_bin}). Any attempt to $t_0$ tag single energy depositions using diffusion would need to tune the simulation to the data with great care. 

